\documentclass[twocolumn,showkeys,showpacs,preprintnumbers,prd,superscriptaddress,nofootinbib]{revtex4-1}

\usepackage{graphicx,epsf,bm,amsmath,amsfonts,amssymb,epstopdf,natbib,hyperref,color,verbatim,multirow,bm,appendix}
\hypersetup{colorlinks=true,urlcolor=blue,citecolor=blue,linkcolor=blue,menucolor=blue,anchorcolor=blue,filecolor=blue}

\date{\today}

\begin{document}

\title{Quasinormal modes-shadow correspondence for rotating regular black holes}

\author{Davide Pedrotti}
\email{davide.pedrotti-1@unitn.it}
\affiliation{Department of Physics, University of Trento, Via Sommarive 14, 38123 Povo (TN), Italy}
\affiliation{Trento Institute for Fundamental Physics and Applications (TIFPA)-INFN, Via Sommarive 14, 38123 Povo (TN), Italy}

\author{Sunny Vagnozzi}
\email{sunny.vagnozzi@unitn.it}
\affiliation{Department of Physics, University of Trento, Via Sommarive 14, 38123 Povo (TN), Italy}
\affiliation{Trento Institute for Fundamental Physics and Applications (TIFPA)-INFN, Via Sommarive 14, 38123 Povo (TN), Italy}

\begin{abstract}
\noindent Eikonal quasinormal modes (QNMs) of black holes (BHs) and parameters of null geodesics, ultimately tied to the appearance of BHs to external observers, are known to be related, and the eikonal QNM-BH shadow radii correspondence has been extensively studied for spherically symmetric BHs. The extension to rotating BHs is non-trivial, and has been worked out only for equatorial ($m=\pm\ell$) QNMs, or for general modes but limited to the Kerr metric. We extend the QNM-shadow radius correspondence to more general rotating space-times, and argue that the requirements for it to hold amount to conditions on the separability of the Hamilton-Jacobi equation for null geodesics and the Klein-Gordon equation. Metrics obtained by the Newman-Janis algorithm enjoy these conditions, provided certain mathematical requirements are imposed on the line element. We explicitly verify the correspondence for the rotating Bardeen and Hayward regular BHs, both of which satisfy the separability requirements. Our findings show that the QNM-shadow radius correspondence holds for a wide range of axisymmetric space-times beyond Kerr. This paves the way to potential strong-field multi-messenger tests of fundamental physics by hearing (via gravitational wave spectroscopy) and seeing (via VLBI imaging) BHs, although substantial improvements relative to the current observational sensitivity are required to make this possible.
\end{abstract}

\maketitle

\section{Introduction}
\label{sec:introduction}

Black holes (BHs) are arguably among the most peculiar objects in the Universe~\cite{Cardoso:2019rvt}. They are commonly believed to be the end product of the evolution of sufficiently massive stars, or more generally the end state of gravitational collapse of matter, and a widespread hope is that they hold the key for the ultimate dream of unifying General Relativity (GR) and Quantum Mechanics (QM). We now know that BHs appear in a broad range of astrophysical environments and come in a very wide range of masses (as large as $10^{10}M_{\odot}$ or larger), in turn leading to a vast array of observational signatures~\cite{Bambi:2017iyh}. Over the past decades, several of these signatures have been observed, providing various direct and indirect indications for the existence of astrophysical BHs. As a result, BHs have gone from being mere academic exercises to providing what is probably one of the most promising stages for observational tests of gravity, and more generally fundamental physics, in the strong-field regime~\cite{Barack:2018yly}.

Despite being well-tested and extremely successful in the weak-field regime, there are many good reasons to believe that GR cannot be the end of the story. These range from the so far unsolved quest for the unification of gravity and QM, to the apparent contradiction between QM unitary evolution and Hawking radiation~\cite{Hawking:1976ra,Harlow:2014yka,Polchinski:2016hrw}, to a wide variety of cosmological and astrophysical observations indicating the existence of unknown dark matter and dark energy components, and finally the fact that continuous gravitational collapse in GR leads to the undesirable appearance of singularities~\cite{Penrose:1964wq,Hawking:1970zqf}. This ``singularity problem'' is probably among the most important issues in fundamental physics, and has motivated the construction of several so-called ``regular'' BH space-times, free of curvature singularities~\cite{Ansoldi:2008jw,Nicolini:2008aj,Sebastiani:2022wbz,Torres:2022twv,Lan:2023cvz}. Nevertheless, its enormous success in the weak-field regime implies that, if GR is to break down, it has to do so in the strong-field regime, e.g.\ in the vicinity of very compact objects. It is for these and other important reasons that the study of strong-field tests of fundamental physics around BHs, boosted by groundbreaking observational and experimental progress, have recently attracted significant attention within the community~\cite{Berti:2015itd}.

Among the wide range of observational signatures of BHs, two are of particular interest to us: the emission of gravitational waves (GWs) from the coalescence of compact binaries (at least one component of which being a BH)~\cite{Cai:2017cbj}, and Very Long Baseline Interferometry (VLBI) horizon-scale images of supermassive BHs (SMBHs)~\cite{Chen:2022scf}. Focusing on the former signature, the final stage of a binary BH merger is the so-called ringdown phase, which is associated with the oscillations of the remnant BH and, except for the late gravitational tail~\cite{Price:1971fb,Price:1972pw}, is characterized by a set of complex quasinormal modes (QNMs). QNMs are the natural, resonant modes of BH perturbations: they are complex, as a consequence of the system in question being dissipative, due to the purely absorbing ability of the event horizon (see e.g.\ Refs.~\cite{Kokkotas:1999bd,Berti:2009kk,Konoplya:2011qq} for reviews). Turning to VLBI BH images, the main features observed therein are a bright emission ring surrounding a central brightness depression~\cite{Synge:1966okc,Luminet:1979nyg,Virbhadra:1999nm,Falcke:1999pj,Narayan:2019imo}. The latter is related to the so-called BH shadow, whose edge marks the apparent image of the photon region (the boundary of the region of space-time supporting closed photon orbits), and separates capture orbits from scattering orbits (see e.g.\ Refs.~\cite{Cunha:2018acu,Dokuchaev:2019jqq,Perlick:2021aok,Wang:2022kvg,AbhishekChowdhuri:2023ekr} for reviews). Around a hundred GW events from mergers involving BHs have so far been detected by the LIGO and Virgo collaborations (with the number expected to increase at a rate of more than one per week once the detectors reach their maximum sensitivity)~\cite{KAGRA:2021vkt}, whereas the groundbreaking images from the Event Horizon Telescope (EHT) collaboration resolved the near-horizon region of the SMBHs M87$^{\star}$~\cite{EventHorizonTelescope:2019dse} and Sagittarius A$^{\star}$ (Sgr A$^{\star}$)~\cite{EventHorizonTelescope:2022wkp} in 2019 and 2022 respectively. Quite literally, we are now able to hear and see BHs, something which would have quite simply been unthinkable until a couple of decades ago.~\footnote{Observations of GWs and BH images have been used to probe and constrain a wide range of fundamental physics scenarios, see e.g.\ Refs.~\cite{Creminelli:2017sry,Sakstein:2017xjx,Ezquiaga:2017ekz,Boran:2017rdn,Baker:2017hug,Amendola:2017orw,Visinelli:2017bny,Crisostomi:2017lbg,Dima:2017pwp,Cai:2018rzd,Oost:2018tcv,Casalino:2018tcd,LIGOScientific:2018dkp,Casalino:2018wnc,Held:2019xde,Bambi:2019tjh,Jusufi:2019nrn,Vagnozzi:2019apd,Zhu:2019ura,Qi:2019zdk,Neves:2019lio,Cunha:2019ikd,Banerjee:2019nnj,Banerjee:2019xds,Kumar:2019pjp,Allahyari:2019jqz,Vagnozzi:2020quf,Lin:2020wnp,Liu:2020ola,Wei:2020ght,Kumar:2020owy,Odintsov:2020sqy,Odintsov:2020zkl,Chen:2020aix,Khodadi:2020jij,Kumar:2020yem,Odintsov:2020xji,Zeng:2020vsj,Odintsov:2020mkz,Khodadi:2020gns,Afrin:2021imp,Eichhorn:2021iwq,Pantig:2021zqe,EventHorizonTelescope:2021dqv,Khodadi:2021gbc,Frion:2021jse,Okyay:2021nnh,Jusufi:2021fek,Guo:2021wid,Roy:2021uye,Chen:2022nbb,Oikonomou:2022ksx,Vagnozzi:2022moj,Ling:2022vrv,Kuang:2022xjp,Guerrero:2022msp,Vagnozzi:2022tba,Eichhorn:2022fcl,Pantig:2022ely,Ghosh:2022kit,Kuang:2022ojj,Khodadi:2022pqh,Banerjee:2022iok,KumarWalia:2022aop,Banerjee:2022bxg,Mustafa:2022xod,Shaikh:2022ivr,Pantig:2022qak,Pantig:2022gih,Odintsov:2022umu,Atamurotov:2022nim,Oikonomou:2022tjm,Sengo:2022jif,Ghosh:2022gka,Afrin:2022ztr,Frizo:2022jyz,Atamurotov:2022knb,Parbin:2022iwt,deLaurentis:2022oqa,Wen:2022hkv,Olmo:2023lil,Karmakar:2023mhs,Pantig:2023yer,Ditta:2023wye,Gonzalez:2023rsd,Sahoo:2023czj,Nozari:2023flq,Gogoi:2023fow,Gogoi:2023ffh,daSilva:2023jxa,Afrin:2023uzo,Arora:2023ijd,Lambiase:2023zeo,Uniyal:2023ahv,Mandal:2023eae,DeMartino:2023ovj,Zubair:2023cor,Belhaj:2023pap,EventHorizonTelescope:2022xqj,Raza:2023vkn,Hoshimov:2023tlz,Chakhchi:2024tzo,Hamil:2024rsg,Chen:2024hpw} for an inevitably incomplete selection of examples in this sense.}

Are QNMs and shadows connected? Let us try and build an intuition for why they should be, starting from the simplest case of spherically symmetric BH space-times. Keeping in mind that QNMs are labeled by three integers $n$, $\ell$, and $m$ (with $n$ being the overtone number, and $\ell$ and $m$ corresponding to the multipolar indices of the QNM angular eigenfunctions), in the eikonal limit ($\ell \gg 1$)~\footnote{This limit allows us to move to the geometric-optics approximation, where the wavelength of the propagating waves is significantly shorter than any other length scale in the system.} we expect high-frequency waves to propagate in a similar manner as photons, leading to the expectation that eikonal QNMs should bear some correspondence to, loosely speaking, parameters characterizing null geodesics. This connection was indeed formalized in a seminal paper by Ferrari and Mashhoon in 1984~\cite{Ferrari:1984zz} (later generalized to stationary, spherically symmetric, asymptotically flat metrics in Ref.~\cite{Cardoso:2008bp}, see also Refs.~\cite{Abramowicz:1997qk,Hod:2009td}), where it was shown that for a Schwarzschild BH, the real part of QNMs in the eikonal limit is proportional to the angular velocity of the last null circular orbit (the photon ring), whereas the imaginary part is related to the Lyapunov exponent, determining the instability time scale of the orbit. Such a connection hints towards an intuitive physical description of QNMs (to the best of our knowledge first suggested by Goebel in 1972~\cite{Goebel:1972ghw}), which can be understood as null particles trapped on the photon ring and diffusing away on a timescale given by the Lyapunov exponent.

From the above discussion, it is clear that unstable null geodesics are intimately tied to QNMs. On the other hand, the optical appearance of BHs is known to be closely linked to the structure of unstable null geodesics. In fact, given that these separate capture orbits from scattering orbits, the so-called BH shadow is none other than the apparent (gravitationally lensed) image of the photon region~\cite{Perlick:2021aok}. These considerations lead to the expectation that there should be a close relation between eikonal QNMs and BH shadows, and more precisely between the real of part of eikonal QNMs and the size of BH shadows, to be defined rigorously later.

To the best of our knowledge, the explicit connection between eikonal QNMs and BH shadows was first made by Jusufi~\cite{Jusufi:2019ltj}, while the important earlier step of explicitly recognizing the close connection between eikonal QNMs and gravitational lensing in the strong deflection limit has been made by Stefanov \textit{et al.}~\cite{Stefanov:2010xz} (see also Refs.~\cite{Andersson:1995vi,Decanini:2002ha,Dolan:2010wr,Khanna:2016yow,Churilova:2019jqx,Wei:2019jve} for important earlier works on the connection between QNMs and critical impact parameter). In particular, in Ref.~\cite{Jusufi:2019ltj} Jusufi argued that for certain classes of static spherically symmetric BH metrics, the real part of QNMs in the eikonal limit is inversely proportional to the shadow radius, with the constant of proportionality being $\ell$ (or, more precisely, $\ell+1/2 \to \ell$ as $\ell \to \infty$). Note that the shadow cast by static spherically symmetric BHs is, by symmetry arguments, a perfect circle, so the BH shadow radius can be defined without ambiguity. The QNM-shadow correspondence has then been studied for several other space-times (see e.g.\ Refs.~\cite{Cuadros-Melgar:2020kqn,Jusufi:2020agr,Hendi:2020knv,Guo:2020nci,Jusufi:2020mmy,Cai:2020kue,Jusufi:2020odz,Jusufi:2020wmp,Mondal:2020pop,Saurabh:2020zqg,Jafarzade:2020ilt,Jafarzade:2020ova,Ghasemi-Nodehi:2020oiz,Cai:2020igv,Campos:2021sff,Cai:2021ele,Li:2021zct,Anacleto:2021qoe,Wu:2021pgf,Liu:2022plm,Konoplya:2022hll,Yu:2022yyv,Lambiase:2023hng,Yan:2023pxj,Das:2023ess,Konoplya:2023moy,Bolokhov:2023dxq,Gogoi:2024vcx,Giataganas:2024hil}), and has been argued to hold quite generically, except in certain special scenarios, including but not limited to the case where photons are non-minimally coupled to other degrees of freedom, resulting in a non-trivial photon propagation which violates the eikonal correspondence~\cite{Konoplya:2017wot}.~\footnote{See for instance Refs.~\cite{Glampedakis:2019dqh,Chen:2019dip,Silva:2019scu,Chen:2021cts,Li:2021mnx,Moura:2021eln,Bryant:2021xdh,Nomura:2021efi,Guo:2021enm,Konoplya:2022gjp} for examples of studies in these directions, and Ref.~\cite{Chen:2022ynz} for a study of the eikonal correspondence in cases with lower degree of symmetry.}

The extension of the above arguments to axisymmetric (rotating) space-times is non-trivial. To begin with, the wave equations and therefore the computation of QNMs becomes significantly more involved. Next, the shadow for rotating BHs is no longer a circle, but is flattened on one side as a result of the interplay between frame dragging effects and the direction of photon spin: this makes the concept of shadow radius, and indirectly the connection to eikonal QNMs (if any), more difficult to establish. Nevertheless, two important works by Jusufi~\cite{Jusufi:2020dhz} and Yang~\cite{Yang:2021zqy} have been developed in this direction. In Ref.~\cite{Jusufi:2020dhz}, Jusufi defined the (typical) shadow radius as the midpoint between the leftmost and rightmost points on the shadow profile in celestial coordinates, linking this radius to the real part of eikonal QNMs. While this definition of shadow radius makes the correspondence applicable to a wide range of metrics beyond the Kerr one,~\footnote{In fact, Jusufi's results are in principle applicable to several axisymmetric space-times, and in Ref.~\cite{Jusufi:2020dhz} the correspondence is explicitly examined and discussed only for the Kerr-Newman and rotating Myers-Perry BHs.} the correspondence itself is intrinsically limited to equatorial photon orbits (since these are the ones that project to the largest apparent radius as seen by a distant observer, and therefore to the leftmost and rightmost points on the shadow boundary), or equivalently to $m=\pm\ell$ QNMs. On the other hand in Ref.~\cite{Yang:2021zqy}, building upon the earlier results of Ref.~\cite{Yang:2012he}, Yang showed how the shadow of a Kerr BH can be mapped to a family of eikonal QNMs: in essence, each and every single point on the shadow boundary can be linked to an eikonal QNM labeled by a certain value of $m/(\ell+1/2)$. The QNM-shadow radius correspondence developed by Yang is more general than the one of Jusufi, given its applicability beyond equatorial QNMs, yet to the best of our knowledge it has only been studied for the Kerr BH case: in fact, essentially all other works exploring the QNM-shadow correspondence for rotating metrics adopt Jusufi's definition of typical shadow radius. While it is not at all unreasonable to expect that Yang's QNM-shadow correspondence should extend beyond the Kerr metric, the extent to which the previous statement is true has yet to be explored.

The above discussion makes it clear that there is a gap in the literature on the QNM-shadow correspondence for rotating space-times~\cite{Jusufi:2020dhz,Yang:2021zqy}, which our work aims to fill. More specifically, our goal is two-fold:
\begin{itemize}
\item to discuss the conditions under which Yang's correspondence holds beyond Kerr BHs;
\item to apply our findings to interesting case studies, specifically rotating regular BHs.
\end{itemize}
With some poetic license, our goal is thus to understand under what conditions we can better understand BHs by seeing the lightning (the shadow) and hearing the thunder (the GW signal) -- or, more likely, the other way around. As we shall see, for the first goal we argue that the requirement amounts to conditions on the separability of the Hamilton-Jacobi equation for null geodesics and Klein-Gordon equation. For the second goal, given the importance of the singularity problem as a motivation for new physics beyond the Einstein-Maxwell system of GR and electromagnetism, we apply our findings to two regular metrics which have received significant attention in the literature, namely the rotating versions of the Bardeen and Hayward BHs~\cite{Bardeen:1968ghw,Hayward:2005gi,Bambi:2013ufa}.

The rest of this paper is then organized as follows: in Section~\ref{sec:kerr} we briefly review the QNM-BH shadow correspondence for the Kerr metric identified by Yang in Ref.~\cite{Yang:2021zqy}. In Section~\ref{sec:regular} we introduce the two rotating BH metrics to which we will apply our subsequent findings, namely the rotating versions of the Bardeen and Hayward regular BHs. In Section~\ref{sec:correspondence} we extend Yang's work to rotating BHs beyond Kerr, discussing the separability conditions under which such an extension holds: we then apply our results to the Bardeen and Hayward rotating regular BHs, numerically verifying the agreement between eikonal QNMs and associated shadow quantities for both metrics. Finally, in Sec.~\ref{sec:conclusions} we draw concluding remarks. A number of technical aspects of our work which may be of interest to some readers but may distract the reading flow of others are instead discussed in Appendices~\ref{sec:appendixa},~\ref{sec:appendixb}, and~\ref{sec:appendixc}. Unless otherwise specified, we use geometrized units with $G=c=1$, label our QNMs by the overtone number $n$, the angular node number $\ell$, and the azimuthal node number $m$, and use the ``mostly plus'' metric signature $(-,+,+,+)$.
 
\section{Quasinormal modes-shadow correspondence for Kerr black holes}
\label{sec:kerr}

\begin{figure*}[!ht]
\centering
\includegraphics[width=0.9\linewidth]{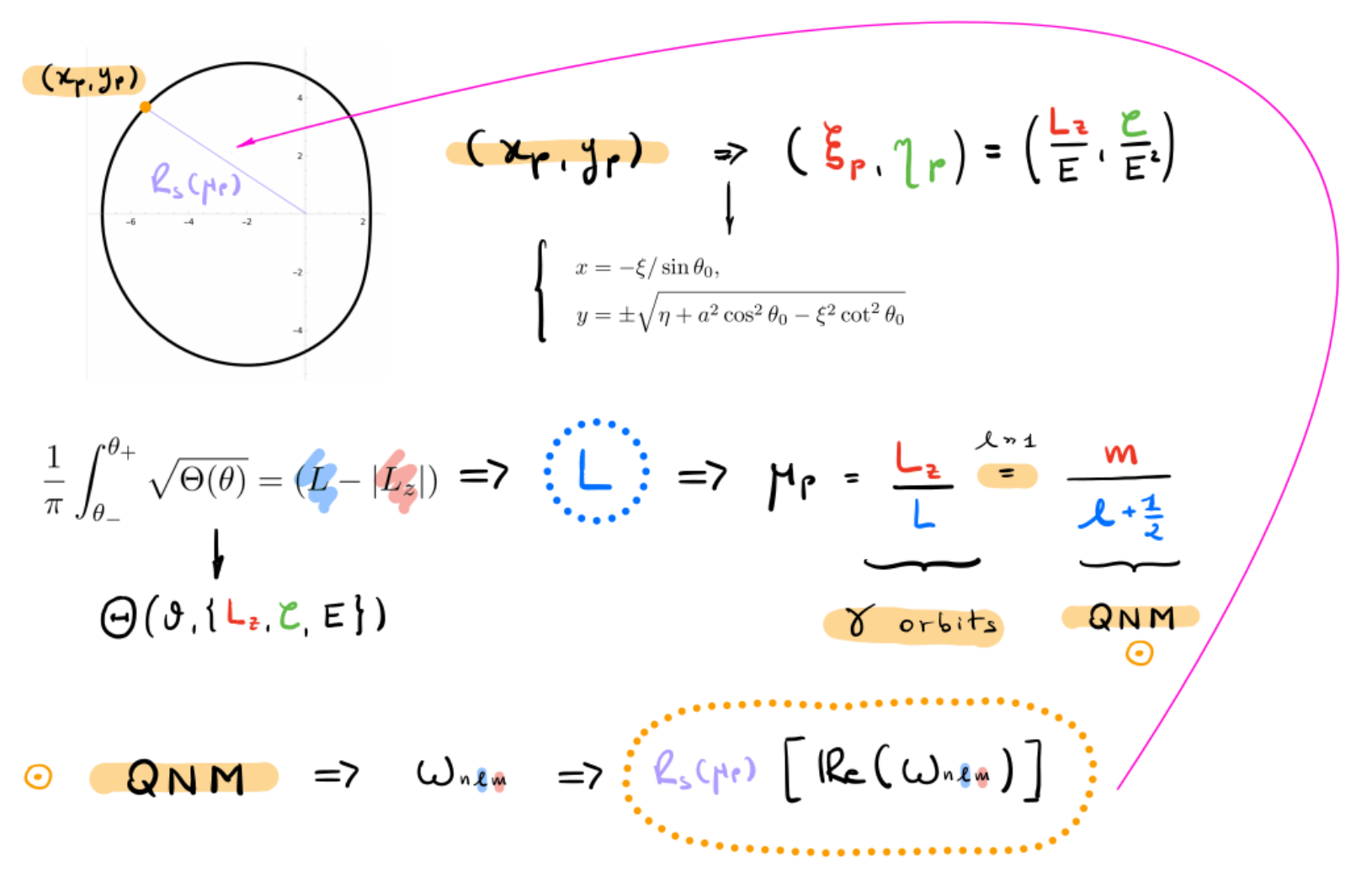}
\caption{Graphical summary of the QNM-shadow correspondence discussed in Sec.~\ref{sec:kerr}.}
\label{fig:graphicalsummary}
\end{figure*}

Black hole quasinormal modes are the natural, resonant modes of BH perturbations or, to put it differently, the characteristic modes of oscillations of BHs, when linearly perturbed either in the metric or with external test fields~\cite{Kokkotas:1999bd,Berti:2009kk,Konoplya:2011qq}. They are characterized by complex frequencies labeled by three indices $n$, $\ell$, and $m$ (whose physical meaning has been discussed in Sec.~\ref{sec:introduction}), $\omega_{n\ell m}=\omega_{n\ell m}^R+i\omega_{n\ell m}^I$, and therefore describe damped oscillatory modes. Mathematically speaking, QNMs correspond to the resonances of the scattering problem to which Sommerfeld boundary conditions, i.e.\ waves which are purely outgoing at infinity and purely ingoing at the event horizon, have been applied. The imaginary part of QNMs, and therefore the damping, reflects the absorbing nature of the event horizon, and is the reason why the modes are \textit{quasinormal}, as opposed to the \textit{normal} modes into which periodic, ever-lasting oscillations are typically decomposed. Henceforth, to not make the notation too heavy, we drop the $_{n\ell m}$ labels (unless explicitly required), and therefore denote $\omega_{n\ell m}^R \to \omega_R$ and $\omega_{n\ell m}^I \to \omega_I$.

As mentioned in Sec.~\ref{sec:introduction}, earlier works in Refs.~\cite{Ferrari:1984zz,Cardoso:2008bp,Hod:2009td,Stefanov:2010xz,Jusufi:2019ltj} recognized the correspondence between unstable geodesics (and thereby ultimately BH shadows) and QNMs in the eikonal limit, i.e.\ for $\ell \gg 1$. This reflects the fact that in the eikonal, geometric-optics approximation, wherein the propagating waves have wavelengths much shorter than any other length scale in the system, the propagation of waves resembles that of photons. Focusing on the QNM-shadow correspondence, it was shown that, for spherically symmetric space-times, the BH shadow radius $R_s$ is related to the real part of eikonal QNMs via~\cite{Jusufi:2019ltj}:
\begin{eqnarray}
R_s = \lim_{\ell \to \infty} \frac{\ell}{\omega_R} = \lim_{\ell \to \infty} \frac{\ell}{\Re(\omega_{n\ell m})}\,.
\label{eq:rsqnm}
\end{eqnarray}
Although not of interest to the rest of this paper, the imaginary part of eikonal QNMs is related to the amplitude ratio between the $N$th and $(N+2)$th photon rings observed in an imaging experiment, which is controlled by the Lyapunov exponent.

The above correspondence holds for a wide class of non-rotating (static, spherically symmetric) BHs. On the other hand, a complete study of the correspondence between QNMs and geodesic quantities for Kerr BHs \textit{for generic values of the BH angular momentum} was, to the best of our knowledge, first examined by Yang \textit{et al.} in Ref.~\cite{Yang:2012he}.~\footnote{Ferrari and Mashhoon~\cite{Ferrari:1984zz} only extended their study to the slowly-rotating case.} In particular, Ref.~\cite{Yang:2012he} found that four parameters characterizing the geometric properties of photon orbits, namely the energy $E$, azimuthal angular momentum $L_z$, Carter constant ${\cal C}$, and Lyapunov exponent $\gamma$ can be mapped one-to-one with eikonal QNMs, as follows~\cite{Yang:2012he}:
\begin{eqnarray}
&&E \leftrightarrow \omega_R\,,\label{eq:mape}\\
&&L_z \leftrightarrow m\,,\label{eq:maplz}\\
&&{\cal C} + L_z^2 \leftrightarrow \Re(A_{\ell m})\,,\label{eq:mapd} \\
&&\gamma \leftrightarrow \omega_I\,,\label{eq:mapgamma}
\end{eqnarray}
where $A_{\ell m}$ is the angular eigenvalue of the QNM~\cite{Teukolsky:1972my}. Moreover, Ref.~\cite{Yang:2012he} succeeded in determining an (approximate) closed expression for eikonal QNMs of Kerr BHs in terms of angular frequency $\Omega_{\theta}$ and precession frequency $\Omega_{\text{prec}}$ of null geodesics as follows:
\begin{eqnarray}
\omega_{\ell m n} \approx \left ( \ell + \frac{1}{2} \right ) \Omega_R(\mu) - i \left ( n + \frac{1}{2} \right )\Omega_I(\mu)
\label{eq:omegaOmega}
\end{eqnarray}
where the quantity $\mu$ is defined as:
\begin{eqnarray}
\mu \equiv \frac{m}{\ell+\frac{1}{2}} \longleftrightarrow \frac{L_z}{L}\,,
\label{eq:mu}
\end{eqnarray}
and can therefore be mapped to $L_z/L$ in the eikonal limit, whereas we have defined:
\begin{eqnarray}
\Omega_R \equiv \Omega_{\theta}(\mu) + \mu\Omega_{\text{prec}}(\mu)\,, \quad \Omega_I \equiv \gamma\,,
\label{eq:omegas}   
\end{eqnarray}
with $\gamma$ being once more the Lyapunov exponent.

Having discussed the above important prerequisites, we now review the arguments put forward by Yang in Ref.~\cite{Yang:2021zqy} to relate eikonal QNMs and the shadow edge for Kerr BHs. Even though this has already been discussed in Ref.~\cite{Yang:2021zqy}, we find it important to revisit the argument, both given its importance for the remainder of our work, and also to clarify a few subtleties which may not be obvious to a reader of Ref.~\cite{Yang:2021zqy}.~\footnote{We strongly encourage the interested reader to consult Ref.~\cite{Yang:2021zqy} for a much more detailed and in-depth discussion on the QNM-shadow radius correspondence.} We recall that Yang's correspondence is more general than the one identified by Jusufi for equatorial photon orbits in Ref.~\cite{Jusufi:2020dhz}, and maps each and every single point on the shadow boundary to an eikonal QNM labeled by a certain value of $\mu \equiv m/(\ell+1/2)$. Null geodesics in Kerr space-time can be described by the Hamilton-Jacobi equation:
\begin{eqnarray}
g^{\mu\nu}\partial_{\mu}S\partial_{\nu}S=0\,,
\label{eq:hamilton}
\end{eqnarray}
where $S(x^{\mu})$ is Hamilton's principal function. As is well-known, the Hamilton-Jacobi equation on Kerr space-time is separable, due to the existence of the Carter constant ${\cal C}$, a quantity which is conserved along geodesic motion together with $E$ and $L_z$. One can then consider a ray moving along a spherical photon orbit, so that the radial component of $S$ is $S_r=0$. For a full cycle along the polar direction labeled by $\theta$, the variation of Hamilton's principal function of course has to vanish:
\begin{eqnarray}
\Delta S = 0 = \oint d\theta \sqrt{\Theta({\cal C},L_z,E)} + L_z\Delta\phi - ET_{\theta}
\label{eq:cycle_int}
\end{eqnarray}
where the symbol $\oint$ denotes integration over a complete cycle, $T_{\theta}$ is the period of motion in the polar direction, related to the angular frequency $\Omega_{\theta}$ via $\Omega_{\theta}=2\pi/T_{\theta}$, $\Delta\phi$ is the change in azimuthal angle after completing a full polar cycle,~\footnote{$\Delta\phi$ is related to the precession angle change $\Delta\phi_{\text{prec}}$ by $ \Delta\phi = \Delta\phi_{\text{prec}} + 2\pi\text{sgn}(L_z)$, reflecting the fact that following a complete polar cycle the azimuthal angle should have changed by $2\pi$ (with the sign of the change depending on the direction of rotation, and hence the sign of $L_z$)~\cite{Yang:2012he}. Finally, the precession frequency $\Omega_{\text{prec}}$ appearing in Eq.~(\ref{eq:omegas}) is given by $\Omega_{\text{prec}} \equiv \Delta\phi_{\text{prec}}/T_{\theta}$.} and the function $\Theta({\cal C},L_z,E)$ is defined as:
\begin{eqnarray}
\Theta({\cal C},L_z,E) = {\cal C} - \cos^2\theta \left ( \frac{L_z^2}{\sin^2\theta} -a^2E^2 \right ) \,.
\label{eq:thetadlze}
\end{eqnarray}
The polar integration cycle can be written as:
\begin{eqnarray}
\oint d\theta \sqrt{\Theta} = 2 \int_{\theta_-}^{\theta_+} d\theta \sqrt{\Theta}\,,
\label{eq:polar}
\end{eqnarray}
where $\theta_{\pm}$ are the two critical angles such that $\Theta(\theta_{\pm})=0$, with $\theta_+>\theta_-$.

Motivated by the earlier Wentzel-Kramers-Brillouin (WKB) analysis of the angular Teukolsky equation performed in Ref.~\cite{Yang:2012he}, the integration cycle in of Eq.~(\ref{eq:polar}) is then rewritten as:
\begin{eqnarray}
\oint d\theta \sqrt{\Theta}=2 \int_{\theta_-}^{\theta_+} d\theta \sqrt{\Theta} = 2\pi(L-\vert L_z \vert)\,.
\label{eq:BScondition}
\end{eqnarray}
The above is reminiscent of a Bohr-Sommerfeld (B-S) quantization condition for a particle moving in a potential given by $\Theta$. Let us pause for a moment to discuss the physical significance of Eq.~(\ref{eq:BScondition}), which may otherwise look counterintuitive. In the high-frequency, eikonal limit,  wavefronts must have an integral number of oscillations in both the polar and azimuthal angular directions. The wave being single-valued implies that there should be $m$ oscillations in the $\phi$ direction, whereas the B-S quantization condition requires there to be $\ell - \vert m \vert +1/2$ oscillations in the $\theta$ direction, in light of the correspondence established by Eqs.~(\ref{eq:mape},\ref{eq:maplz},\ref{eq:mapd},\ref{eq:mapgamma}). Therefore, the B-S quantization condition implies that the azimuthal angular momentum $L_z$ \textit{and} the Carter constant ${\cal C}$ (or, more precisely, ${\cal C}+L_z^2$) should be quantized so as to obtain a standing wave in both the polar and azimuthal angular directions (see Tab.~I in Ref.~\cite{Yang:2012he}).

Before moving on, let us note that ${\cal C}$, $L_z$, and $E$ are related to two quantities typically denoted by $\xi$ and $\eta$:
\begin{eqnarray}
\xi \equiv \frac{L_z}{E}\,, \quad \eta \equiv \frac{{\cal C}}{E^2}\,,
\label{eq:xieta}
\end{eqnarray}
both of which are obviously constants of motion. It is convenient to work with $\xi$ and $\eta$ in first place because photon trajectories are independent of energy $E$ (as a consequence of the equivalence principle), and second because these variables are directly related to the celestial coordinates of an observer at infinity~\cite{Perlick:2021aok}:
\begin{eqnarray}
x = -\frac{\xi}{\sin\theta_O}\,, \quad y=\pm\sqrt{\eta+a^2\cos^2\theta_O-\xi^2\cot^2\theta_O}\,,
\label{eq:celestialcoordinatesxy}
\end{eqnarray}
where $a$ is the Kerr BH spin parameter, and $\theta_O$ is the angle the observer makes with respect to the BH axis of rotation, with $\theta_O=\pi/2$ denoting an observer on the equatorial plane (``edge-on'').

Let us now return to Eq.~(\ref{eq:BScondition}), defining the B-S quantization condition. Solutions to this equation are not known in closed form, but approximated ones can be obtained as a truncated series expansion as follows~\cite{Yang:2021zqy}:
\begin{eqnarray}
{\cal C} + L_z^2 = L^2 - \frac{a^2E^2}{2} \left ( 1 - \frac{L_z^2}{L^2} \right ) + {\cal O} \left ( \frac{a^4E^4}{L^4} \right ) \,.
\label{eq:expansion}
\end{eqnarray}
To identify the connection with BH shadows, we note that a null ray which just manages to escape to infinity has an impact parameter~\cite{Yang:2021zqy}:
\begin{eqnarray}
R_s(\mu) = \frac{1}{E}\sqrt{{\cal C}+L_z^2}\,,
\label{eq:rs}
\end{eqnarray}
where we have made the dependence on $L_z$, and thereby $\mu=L_z/L$, explicit. Since such a null ray defines the edge of the BH shadow, $R_s$ given in Eq.~(\ref{eq:rs}) can be directly interpreted as being the shadow radius. Note however that, unlike in the case of a Schwarzschild BH where the shadow is a perfect circle, the shadow of a Kerr BH is flattened on one side, and is not symmetric upon reflection around the $y$ axis of a celestial observer at infinity. Therefore, $R_s$ varies from point to point on the shadow edge, and its point-by-point physical interpretation is nothing other than the distance of each point along the shadow edge from the origin of the celestial coordinate system $(x,y)=(0,0)$.~\footnote{Note that for axisymmetric metrics such as the Kerr metric, the celestial origin is no longer the actual geometrical center of the shadow, due to frame-dragging effects which cause a horizontal ``drift'' of the shadow, see e.g.\ Ref.~\cite{Chen:2022nbb}.} Considering that in the eikonal limit $\mu \equiv m/(\ell + 1/2) \approx L_z/L$ as per Eq.~(\ref{eq:mu}), and with $R_s$ given by Eq.~(\ref{eq:rs}), we can now express Eq.~(\ref{eq:expansion}) as:
\begin{eqnarray}
\frac{L^2}{E^2} \approx \frac{\widetilde{L}^2}{E^2} =  R_s(\mu)^2 + \frac{a^2}{2} \left ( 1 - \mu^2 \right ) \,.
\label{eq:expansionnew}
\end{eqnarray}
where we use the notation $\widetilde{L}$ to reflect the fact that we have neglected terms of order $a^4E^4/L^4$ or higher. Therefore, $\widetilde{L}$ is actually only an approximation to $L$, albeit a sufficiently good one for percent-level tests of GR, as argued in Ref.~\cite{Yang:2021zqy}. Finally, combining Eqs.~(\ref{eq:mape},\ref{eq:mu},\ref{eq:maplz},\ref{eq:omegas},\ref{eq:cycle_int},\ref{eq:BScondition}) and recalling that the angular and precession frequencies are $\Omega_{\theta}=2\pi/T_{\theta}$ and $\Omega_{\text{prec}} = \Delta\phi_{\text{prec}}/T_{\theta}$, we obtain:
\begin{eqnarray}
\frac{L}{E} &=& -\frac{L_z\Delta\phi_{\text{prec}}}{2\pi E}+\frac{T_{\theta}}{2\pi} = -\frac{L_z}{E}\frac{\Omega_{\text{prec}}}{\Omega_{\theta}}+\frac{1}{\Omega_{\theta}} \nonumber \\
&=& \frac{1}{\Omega_{\theta}} \left ( -\frac{L_z}{E}\Omega_{\text{prec}} +1 \right ) \to \frac{1}{\Omega_{\theta}} \left ( -\frac{m}{\omega_R}\Omega_{\text{prec}} +1 \right ) \nonumber \\
&=& \frac{1}{\Omega_{\theta}} \left ( -\frac{\mu \left ( \ell+1/2 \right ) }{ \left ( \ell+1/2 \right ) \Omega_R}\Omega_{\text{prec}} +1 \right ) = \frac{1}{\Omega_{\theta}} \left ( -\frac{\mu\Omega_{\text{prec}}}{\Omega_R} + 1 \right ) \nonumber \\
&=& \frac{1}{\Omega_{\theta}} \left ( \frac{\Omega_R-\mu\Omega_{\text{prec}}}{\Omega_R} \right ) = \frac{1}{\Omega_{\theta}}\frac{\Omega_{\theta}}{\Omega_R} = \frac{1}{\Omega_R}\,,
\label{eq:connection}
\end{eqnarray}
where in the fifth equality we have used the fact that $\Omega_R$ is the real part of eikonal QNMs up to a factor of $(\ell+1/2)$. Finally, combining Eqs.~(\ref{eq:expansionnew},\ref{eq:connection}) and approximating $L \approx \widetilde{L}$ we find:
\begin{eqnarray}
R_s(\mu) &\approx& \sqrt{\frac{L^2}{E^2}-\frac{a^2}{2} \left ( 1 - \mu^2 \right ) } = \sqrt{\frac{1}{\Omega_R^2}-\frac{a^2}{2} \left ( 1 - \mu^2 \right ) } \nonumber \\
&\approx& \sqrt{ \left ( \frac{\ell+\frac{1}{2}}{\Re(\omega_{n\ell m})} \right ) ^2-\frac{a^2}{2} \left ( 1 - \mu^2 \right ) } \nonumber \\
&\approx& \sqrt{ \left ( \frac{\ell}{\Re(\omega_{n\ell m})} \right ) ^2-\frac{a^2}{2} \left ( 1 - \mu^2 \right ) } \,,
\label{eq:correspondence}
\end{eqnarray}
where the last two approximations holds in the eikonal limit. Eq.~(\ref{eq:correspondence}) directly establishes a point-by-point connection between the real part of eikonal QNMs [$\Re(\omega_{n\ell m})$ on the right-hand side] and points on the edge of the shadow of Kerr BHs [$R_s(\mu)$ on the left-hand side]. Note that Eq.~(\ref{eq:correspondence}) reduces to Eq.~(\ref{eq:rsqnm}) in the non-rotating limit $a \to 0$, recovering the well-known correspondence for static spherically symmetric metrics.

Let us now pause and reflect on the meaning of Eq.~(\ref{eq:correspondence}). It might not be immediately obvious to the reader how a value of $\mu$ can be associated to each point on the edge of the BH shadow. The fact that each point on the edge of the shadow can be linked to an eikonal QNM labeled by a certain value of $\mu=m/(\ell+1/2)$ is indeed what the left-hand side of Eq.~(\ref{eq:correspondence}) implies, but the map ``point$\,\,\longleftrightarrow \mu$'' is not obvious. Let us start by considering a point $P$ on the edge of the BH shadow, labeled by celestial coordinates $(x_P,y_P)$, and at a distance $R_s$ from the celestial origin. Inverting Eq.~(\ref{eq:celestialcoordinatesxy}) we can obtain the values of the conserved quantities $(\xi_P,\eta_P) \equiv ((L_z/E)_P,({\cal C}/E^2)_P)$ associated to the photon orbit whose projection on the sky of the observer corresponds to the specific point $P$. Through the B-S quantization condition we can then obtain the angular momentum $L$, or more precisely $L/E$. To see this, note that Eq.~(\ref{eq:BScondition}), once combined with Eq.~(\ref{eq:thetadlze}), is an equation determining $L/E$ as a function of the two constants of motion $\eta \equiv {\cal C}/E^2$ and $\xi \equiv L_z/E$, rather than ${\cal C}$, $L_z$, and $E$ independently. To make this explicit, we can rewrite Eq.~(\ref{eq:BScondition}) as:
\begin{eqnarray}
&&2 \int_{\theta_-}^{\theta_+} d\theta \sqrt{\frac{{\cal C}}{E^2} - \tan^2\theta \left ( \frac{L_z^2}{E^2} - \frac{a^2}{\sin^2\theta} \right ) } \nonumber \\
&& = 2\pi \left ( \frac{L}{E}- \frac{\vert L_z \vert}{E} \right )
\label{eq:BSconditionDE2LzE}
\end{eqnarray}
With $L_z/E$ and $L/E$ known, through Eq.~(\ref{eq:mu}) we can then determine $\mu=(L_z/E)/(L/E)$. This establishes the map between each point $P$, and a value of $\mu$. To proceed to the QNM part of the correspondence, we note that the value of $\mu$ we have now identified is given (in the eikonal limit) by $\mu \approx m/\ell$, from which we can determine $\ell$ and hence (numerically) $\omega_{n\ell m}$,~\footnote{We recall that $n$ is the overtone number, with higher $n$ modes associated to faster damping. In the eikonal limit ($\ell \gg 1$), the real part of $\omega_{n\ell m}$ is typically independent from $n$ (see \cite{Berti:2009kk} for some examples), which is why the overtone number does not enter in the previous discussion. In practice, in our later numerical study we adopt the Prony method to extract QNM frequencies from the time evolution of the perturbation: this method is essentially only sensitive to the fundamental mode ($n=0$) and the first overtone ($n=1$), which is why we will only consider $n=0$ in what follows.} and therefore $\Re(\omega_{n\ell m})$ as in the right-hand side of Eq.~(\ref{eq:correspondence}). The value of $R_s(\mu)$ computed through Eq.~(\ref{eq:correspondence}) then matches the distance between the point $P$ and the celestial origin. This concludes the QNM-shadow correspondence for Kerr BHs, which is also graphically summarized in Fig.~\ref{fig:graphicalsummary}. Five comments are in order before we move on:
\begin{itemize}
\item the ``point$\,\,\longleftrightarrow \mu$'' map is not a linear one, and has to be established numerically due to the absence of closed solutions to the B-S quantization condition;
\item the observer's inclination angle sets the range of allowed values of $\mu$, which in any case is necessarily bounded by $-1 \leq \mu_{\min} \leq \mu \leq \mu_{\max} \leq 1$;
\item the points on the edge of the BH shadow corresponding to $\mu_{\min}$ and $\mu_{\max}$ are the leftmost and rightmost points on the celestial $x$ axis respectively (these are not symmetric with respect to the origin), themselves corresponding to equatorial photon orbits,~\footnote{It is to these two points on the BH shadow edge that the work of Ref.~\cite{Jusufi:2020dhz} applies.} whereas $\mu=0$ corresponds to the two points on the celestial $y$ axis (which on the other hand are symmetric with respect to the origin);
\item although in principle $\mu$ is a discrete parameter, in the eikonal limit $\ell \to \infty$ it asymptotically behaves as a continuous parameter (recall that $m=-\ell\,,-\ell+1\,,...,\ell-1\,,\ell$), and in this sense one can in principle reconstruct each point on the BH shadow edge simply by arbitrarily increasing $\ell$;
\item the correspondence described above has only been explicitly tested for the Kerr metric in Ref.~\cite{Yang:2021zqy} -- in fact, one of our goals in this work is to test the correspondence in the context of two well-motivated rotating regular BHs.
\end{itemize}
It is worth stressing that, as discussed in Ref.~\cite{Yang:2021zqy}, the connection we have reviewed above is only an approximate one, since we have approximated $L \approx \widetilde{L}$.

While, as argued by Yang~\cite{Yang:2021zqy}, the approximation $L \approx \widetilde{L}$ is sufficiently accurate for percent-level tests of GR, we note that there is a simple way to make the QNM-shadow correspondence even more accurate beyond what has been done in Ref.~\cite{Yang:2021zqy}. Specifically, let us denote by $\varepsilon(\mu,a)$ the following difference:
\begin{eqnarray}
\frac{\widetilde{L}}{E} \equiv \frac{L}{E} - \varepsilon(\mu,a) = \frac{1}{\Omega_R} - \varepsilon(\mu,a)\,,
\label{eq:vareps}
\end{eqnarray}
which quantifies the error in the approximation given by Eq.~(\ref{eq:expansion}). Then, using Eqs.~(\ref{eq:expansion}--\ref{eq:connection}), we find that the shadow radius can be expressed as follows:
\begin{eqnarray}
R_s(\mu) &=& \sqrt{\left(\frac{1}{\Omega_R} - \varepsilon(\mu,a)\right)^2 - \frac{a^2}{2}(1 - \mu^2)} \nonumber \\
&=& \sqrt{\left( \frac{\ell + 1/2}{\Re(\omega_{n\ell m})} - \varepsilon(\mu,a)\right)^2 - \frac{a^2}{2}(1 - \mu^2)}\,, \nonumber \\
\label{eq:correspondencenew}
\end{eqnarray}
which, we stress, is an equality and not an approximation. When we later test the QNM-shadow correspondence on two rotating regular BH space-times in Sec.~\ref{sec:regular}, the use of Eq.~(\ref{eq:correspondencenew}) and how the various components thereof are computed will become clearer. For the moment let us anticipate that, in doing so, we compute our QNMs $\omega_{n\ell m}$ via a $2+1$ time-evolution code described in Appendix~\ref{sec:appendixa}, while the correction term $\varepsilon(\mu,a)$ is obtained by using a photon orbit code described in Appendix~\ref{sec:appendixc}. The value of $R_s(\mu)$ on the left-hand side can finally be compared with the same quantity calculated numerically using closed photon orbits, with the ``point$\,\,\longleftrightarrow \mu$'' map established numerically.

\section{Rotating regular black holes}
\label{sec:regular}

Continuous gravitational collapse in GR leads to the inevitable existence of singularities, as established by the celebrated Penrose-Hawking singularity theorems~\cite{Penrose:1964wq,Hawking:1970zqf} (see also Refs.~\cite{Hawking:1976ra,Senovilla:1998oua,Senovilla:2014gza}). These (essential) singularities encapsulate two aspects: the divergence of curvature invariants (sets of independent scalars constructed from the Riemann tensor and the metric, e.g.\ $R \equiv g^{\mu\nu}R_{\mu\nu}$, $R_{\mu\nu}R^{\mu\nu}$, and ${\cal K} \equiv R_{\mu\nu\rho\sigma}R^{\mu\nu\rho\sigma}$), as well as the incompleteness of null and timelike geodesics, for which the affine parameter of a test particle terminates at a finite value.~\footnote{We note that in the literature the criterion for regularity is usually limited to the finiteness of curvature invariants -- this does not necessarily imply geodesic completeness (whose inclusion therefore makes the criteria for regularity significantly more stringent), and viceversa.} However, these singularities are arguably somewhat undesirable, as they lead, among other things, to the breakdown of predictability in gravitational collapse (see, however, Refs.~\cite{Sachs:2021mcu,Ashtekar:2021dab,Ashtekar:2022oyq} for a different viewpoint). This is a key motivation for research into mechanisms for singularity resolution which, among other things, might elegantly solve the information paradox~\cite{Hawking:1976ra,Harlow:2014yka,Polchinski:2016hrw}.

Regular BHs (RBHs), or more generally regular space-times, are free of essential singularities in the entire space-time: that is, their curvature invariants are finite everywhere \textit{and} null and timelike geodesics thereon are complete, although this does not of course preclude the presence of coordinate singularities, e.g.\ at the horizon(s). The study of RBHs has a long and rich history, dating back at the least to the works of Sakharov and Gliner in 1966~\cite{Sakharov:1966aja,Gliner:1966ghw}, and later in the 1970s by Gliner, Dymnikova, Gurevich, and Starobinsky~\cite{Gliner:1975ghw,Gurevich:1975ghw,Starobinsky:1979ty}. It is a common belief, although one supported by only a handful of first-principles investigations (see e.g.\ Refs.~\cite{Dymnikova:1992ux,Dymnikova:2004qg,Ashtekar:2005cj,Bebronne:2009mz,Modesto:2010uh,Spallucci:2011rn,Perez:2014xca,Colleaux:2017ibe,Nicolini:2019irw,Bosma:2019aiu,Jusufi:2022cfw,Olmo:2022cui,Jusufi:2022rbt,Ashtekar:2023cod,Nicolini:2023hub}), that quantum gravitational effects on sufficiently small scales (which however could propagate to larger scales and therefore lead to observable effects) could prevent the formation of singularities, although both the actual mechanism and the quantum theory of gravity behind remain as of yet unknown~\cite{Addazi:2021xuf}. As a result, there are a number of (semi)phenomenologically-motivated proposals for RBHs, several of which replace the singular region by a patch of regular space (e.g.\ de Sitter or Minkowski space). Although we cannot do justice to the enormous literature on RBHs, we mention Refs.~\cite{Borde:1996df,AyonBeato:1998ub,AyonBeato:1999rg,Bronnikov:2005gm,Berej:2006cc,Bronnikov:2012ch,Rinaldi:2012vy,Stuchlik:2014qja,Schee:2015nua,Johannsen:2015pca,Myrzakulov:2015kda,Fan:2016hvf,Sebastiani:2016ras,Toshmatov:2017zpr,Chinaglia:2017uqd,Frolov:2017dwy,Bertipagani:2020awe,Nashed:2021pah,Simpson:2021dyo,Franzin:2022iai,Chataignier:2022yic,Khodadi:2022dyi,Sajadi:2023ybm,Javed:2024wbc,Ditta:2024jrv,Al-Badawi:2024lvc,Corona:2024gth,Bueno:2024dgm,Calza:2024fzo,Calza:2024xdh} as examples of proposals for RBHs, whereas we invite the reader to consult Refs.~\cite{Ansoldi:2008jw,Nicolini:2008aj,Sebastiani:2022wbz,Torres:2022twv,Lan:2023cvz} for recent reviews on the subject (see also Ref.~\cite{Bambi:2023try}).~\footnote{An important recent debate in the literature revolves around the issue of whether or not RBHs are stable, see e.g.\ Refs.~\cite{Carballo-Rubio:2018pmi,Bonanno:2020fgp,Carballo-Rubio:2021bpr,Carballo-Rubio:2022kad,Franzin:2022wai,Bonanno:2022jjp,Bonanno:2022rvo,Carballo-Rubio:2022twq,Bonanno:2023qhp,Carballo-Rubio:2024dca} for discussions on the topic. In fact, several RBH metrics feature an (inner) Cauchy horizon whose surface gravity is typically non-zero~\cite{Carballo-Rubio:2019fnb,Bonanno:2020fgp}. This makes the space-time potentially prone to the mass inflation instability~\cite{Poisson:1989zz,Poisson:1990eh,Balbinot:1993rf,Hamilton:2008zz,Barcelo:2022gii}, i.e.\ the exponential growth of the mass function at the Cauchy horizon, ultimately generating a null singularity.}

In this work, we shall be interested in two among the most widely studied RBH metrics: the Bardeen BH and the Hayward BH. While both were originally introduced on purely phenomenological grounds, these two metrics are now considered among the prototypes for RBHs. Note that both the Bardeen and Hayward RBHs are regular in the sense of having finite curvature invariants (more specifically, singularities in the curvature invariants are moved to the non-physical domain, e.g.\ for imaginary values of the radial coordinate), but have been shown to not satisfy the more stringent requirement of geodesic completeness, see e.g.\ Ref.~\cite{Zhou:2022yio}. Nevertheless, we shall consider them mostly for proof-of-principle/case study purposes, especially given the widespread interest in these metrics and their being considered prototypes for RBHs in the recent literature.

In his seminal work of Ref.~\cite{Bardeen:1968ghw}, Bardeen phenomenologically proposed to cure the essential singularity present in the Schwarzschild space-time at $r=0$, by replacing the mass $M$ with an $r$-dependent function $m(r)$. The line element of the Bardeen BH takes the following form:
\begin{eqnarray}
ds^2 = -f(r) dt^2 + f(r)^{-1} dr^2 + r^2d\Omega^2\,,
\label{eq:bardeen}
\end{eqnarray}
where $d\Omega^2$ is the metric on the two-sphere, and the function $f(r)$ is given by~\cite{Bardeen:1968ghw}:
\begin{eqnarray}
f(r) = 1-\frac{2Mr^2}{(r^2+q_m^2)^{3/2}}\,,
\label{eq:bardeenfunction}
\end{eqnarray}
where $q_m$ is a regularization parameter, and the Schwarzschild BH is recovered in the limit $q_m \to 0$. The curvature invariants of the above space-time are finite everywhere, and the BH can be understood to possess a de Sitter core at its center, in place of the would-be singularity. It has been argued that the parameter $q_m$, introduced on purely phenomenological grounds and required to satisfy $q_m/M \leq \sqrt{16/27}$ in order for the metric to possess an event horizon, can be interpreted as being the magnetic charge of a magnetic monopole sourcing the Bardeen space-time~\cite{Ayon-Beato:2000mjt}, potentially in the context of a theory of non-linear electrodynamics~\cite{Ayon-Beato:2004ywd}. Another potential top-down interpretation of the Bardeen space-time is as a quantum-corrected BH in the presence of a generalized uncertainty principle~\cite{Maluf:2018ksj}.

Another well-known RBH was proposed by Hayward in Ref.~\cite{Hayward:2005gi}. The Hayward space-time is controlled by the regularization parameter $\ell$, and also possesses a de Sitter core at its center, parametrized in terms of an effective cosmological constant $\Lambda = 3/\ell^2$. The Hubble length associated to the de Sitter core is required to satisfy $\ell/M \leq \sqrt{16/27}$ for there to be an event horizon. The metric of the Hayward BH can also be written in the form of Eq.~(\ref{eq:bardeen}), with $f(r)$ given by~\cite{Hayward:2005gi}:
\begin{eqnarray}
f(r) = 1-\frac{2Mr^2}{r^3+2M\ell^2}\,.
\label{eq:haywardfunction}
\end{eqnarray}
Much as the Bardeen BH, the Hayward BH has been proposed on purely phenomenological grounds, although it can potentially emerge in the context of the equation of state of matter at high density~\cite{Sakharov:1966aja,Gliner:1966ghw}, within finite density or finite curvature theoretical proposals~\cite{1982JETPL..36..265M,1987JETPL..46..431M,Mukhanov:1991zn} (the latter expected to emerge within a quantum theory of gravity~\cite{AlvesBatista:2023wqm}), and finally within theories of non-linear electrodynamics~\cite{Kumar:2020bqf,Kruglov:2021yya}, as is the case for several RBH metrics~\cite{Bronnikov:2017sgg,Ghosh:2021clx,Bronnikov:2021uta,Bronnikov:2022ofk,Bokulic:2023afx}. 

\begin{figure}[!ht]
\centering
\includegraphics[width=1.0\linewidth]{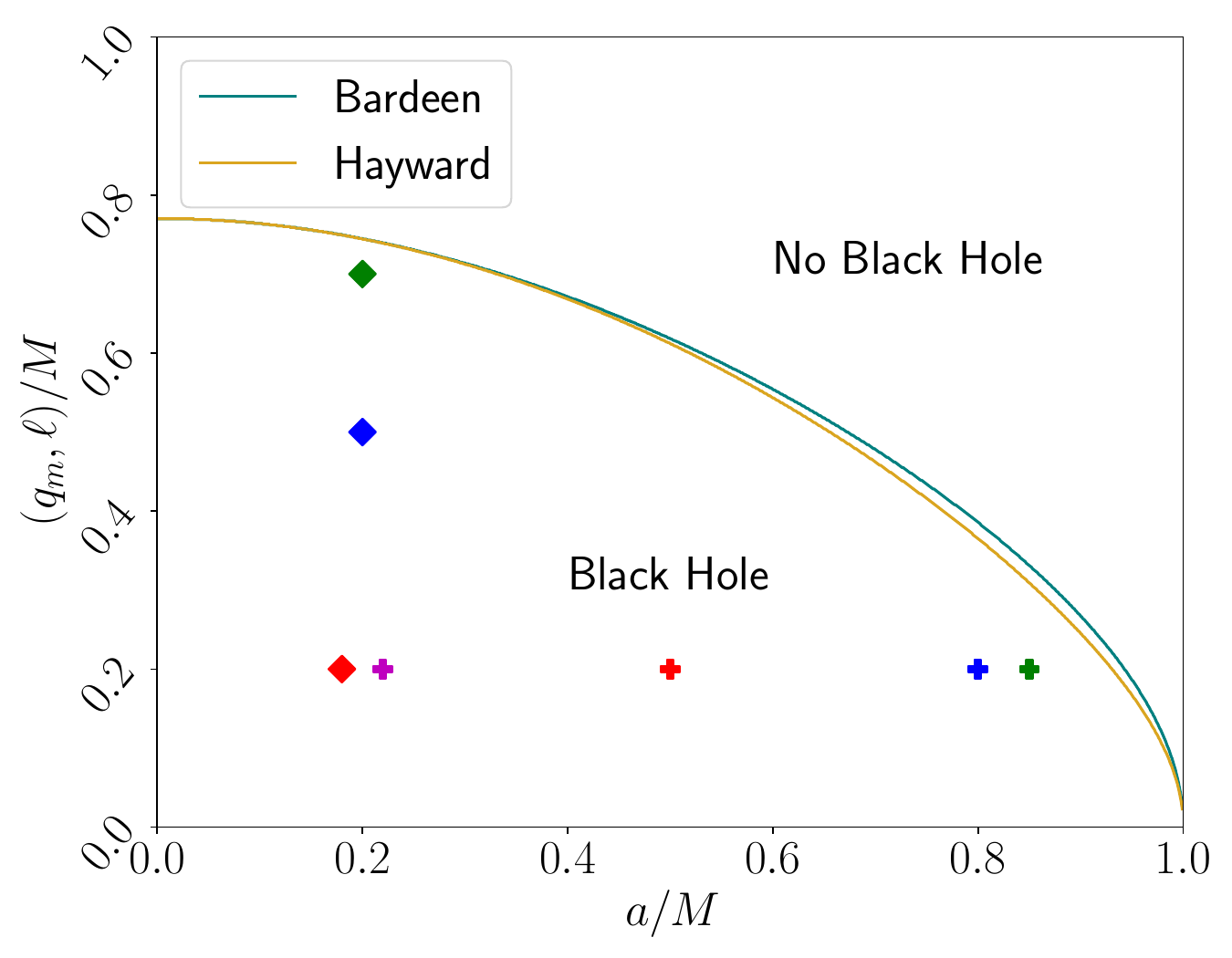}
\caption{Phase diagram of the allowed values of the regularization/hair parameters $q_m$ and $\ell$, for the rotating Bardeen (red curve) and Hayward (blue curve) BHs, as a function of the BH spin parameter $a$: the regions above the curves are not allowed as they would lead to horizonless objects. \textbf{The markers denote the points in parameter space we later sample when explicitly testing the eikonal QNM-shadow correspondence. Diamond markers indicate points at fixed spin and increasing hair parameter, whereas plus sign markers indicate points at fixed hair parameter and increasing spin (the red diamond and magenta plus sign markers are both at $a/M=0.2$ and $\{q_m/M\,,\ell/M\}=0.2$, but have been slightly separated in order to avoid the markers sitting on top of each other)}.}
\label{fig:bardeen_hayward_bh_nobh}
\end{figure}

In the present work, we are interested in the rotating versions of the Bardeen and Hayward space-times described earlier. These were first obtained by Bambi and Modesto in Ref.~\cite{Bambi:2013ufa} making use of the Newman-Janis algorithm~\cite{Newman:1965my,Newman:1965tw}, a five-step method to build an axially symmetric spacetime starting from a spherically symmetric one.~\footnote{See Ref.~\cite{Erbin:2016lzq} for a recent review on the algorithm, and Refs.~\cite{Drake:1998gf,Rajan:2015ffs,Rajan:2016zmq,Arkani-Hamed:2019ymq,Guevara:2020xjx,Mazza:2021rgq,Kamenshchik:2023woo} for further insights on the method and its limitations. In light of ambiguities surrounding the complexification procedure in the Newman-Janis algorithm, we have also adopted the method without complexification proposed in Refs.~\cite{Azreg-Ainou:2014pra,Azreg-Ainou:2014aqa}, verifying that the same metrics are recovered.} In Boyer-Lindquist coordinates, the line elements of the rotating Bardeen and Hayward space-times can be written in the form:
\begin{eqnarray}
ds^2 &=& - \left ( 1 - \frac{2f}{\Sigma} \right ) dt^2 + \frac{\Sigma}{\Delta} dr^2 - \frac{4af\sin^2\theta}{\Sigma}d\phi dt \nonumber \\
&&+\Sigma d\theta^2 + \frac{\sigma\sin^2\theta}{\Sigma}d\phi^2
\label{eq:rotatingbh}
\end{eqnarray}
where we have defined:
\begin{eqnarray}
\Sigma(r,\theta) &\equiv& r^2 + a^2\cos^2\theta\,,\\
\label{eq:sigmacapt}
2f(r) &\equiv& r^2(1 - F)\,,\\
\label{eq:f}
\Delta(r) &\equiv& r^2F +a^2 = r^2 -2f + a^2\,,\\
\label{eq:delta}
\sigma(r,\theta) &\equiv& (r^2 + a^2)^2 -a^2\Delta \sin^2\theta\,,
\label{eq:sigma}
\end{eqnarray}
and the function $F(r)$ for the Bardeen and Hayward BHs ($F_B$ and $F_H$ respectively) takes the form:
\begin{eqnarray}
F_B(r) &\equiv& 1 - \frac{2Mr^2}{(r^2 + q_m^2)^{3/2}}\,, \label{eq:fb} \\
F_H(r) &\equiv& 1 - \frac{2Mr^2}{r^3 + 2M\ell^2} \label{eq:fh} \,.
\end{eqnarray}
The location of the event horizon is determined by setting $\Delta = 0$. In turn, this sets constraints on the allowed values of ($a$, $g$) and ($a$, $\ell$), in order for the space-time to not be horizonless. These constraints are shown in Fig.~\ref{fig:bardeen_hayward_bh_nobh} for the rotating Bardeen (red curve) and Hayward (blue curve) BHs, and will be implicitly imposed in what follows. Note that in the limit of no rotation ($a=0$), these constraint reduce to the well-known limit $q_m,\,\ell \leq \sqrt{16/27}M \approx 0.77M$.

\section{Quasinormal modes-shadow correspondence for rotating regular black holes}
\label{sec:correspondence}

\begin{figure*}[!ht]
\includegraphics[width=0.49\linewidth]{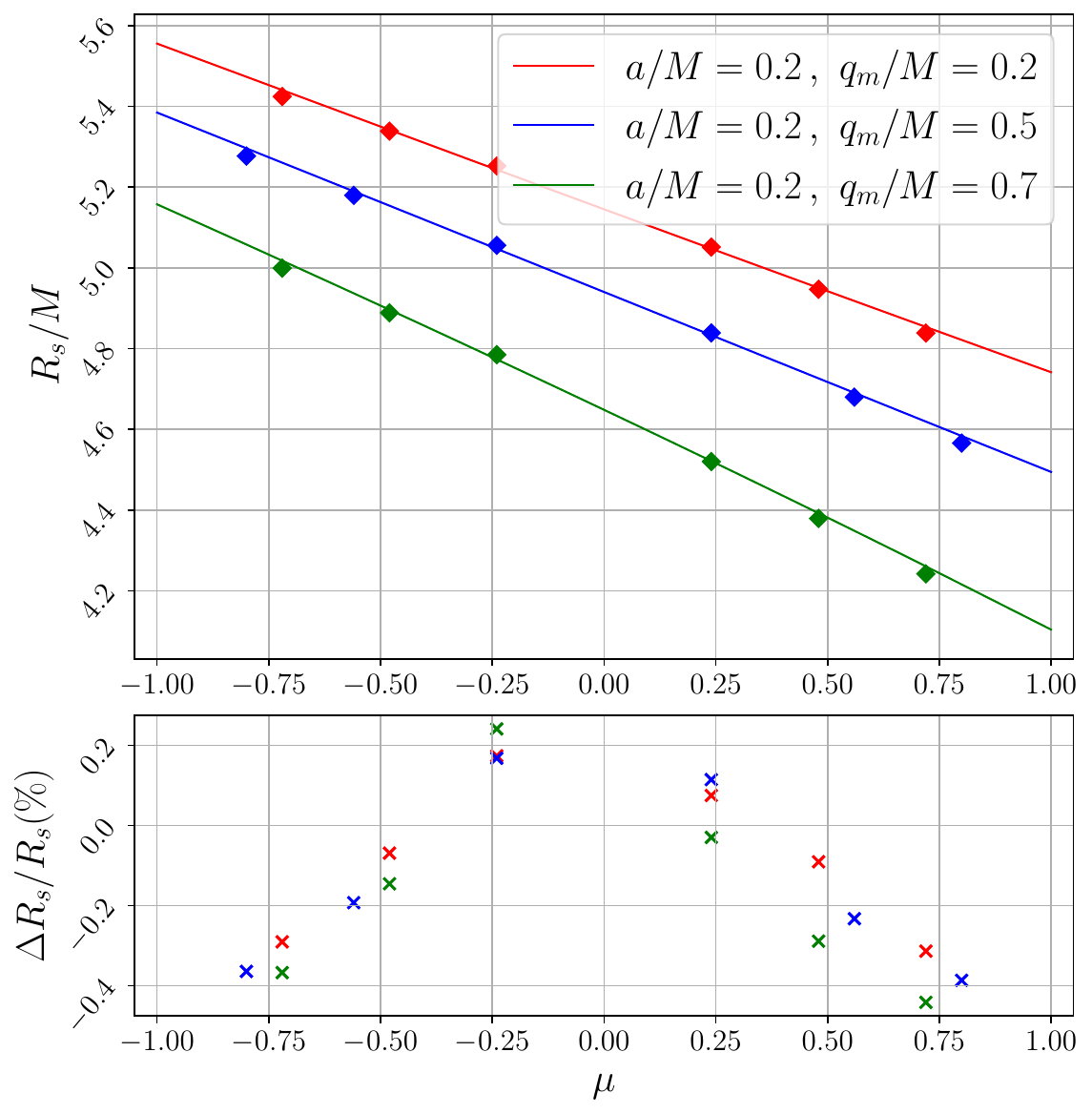}\,\,\includegraphics[width=0.49\linewidth]{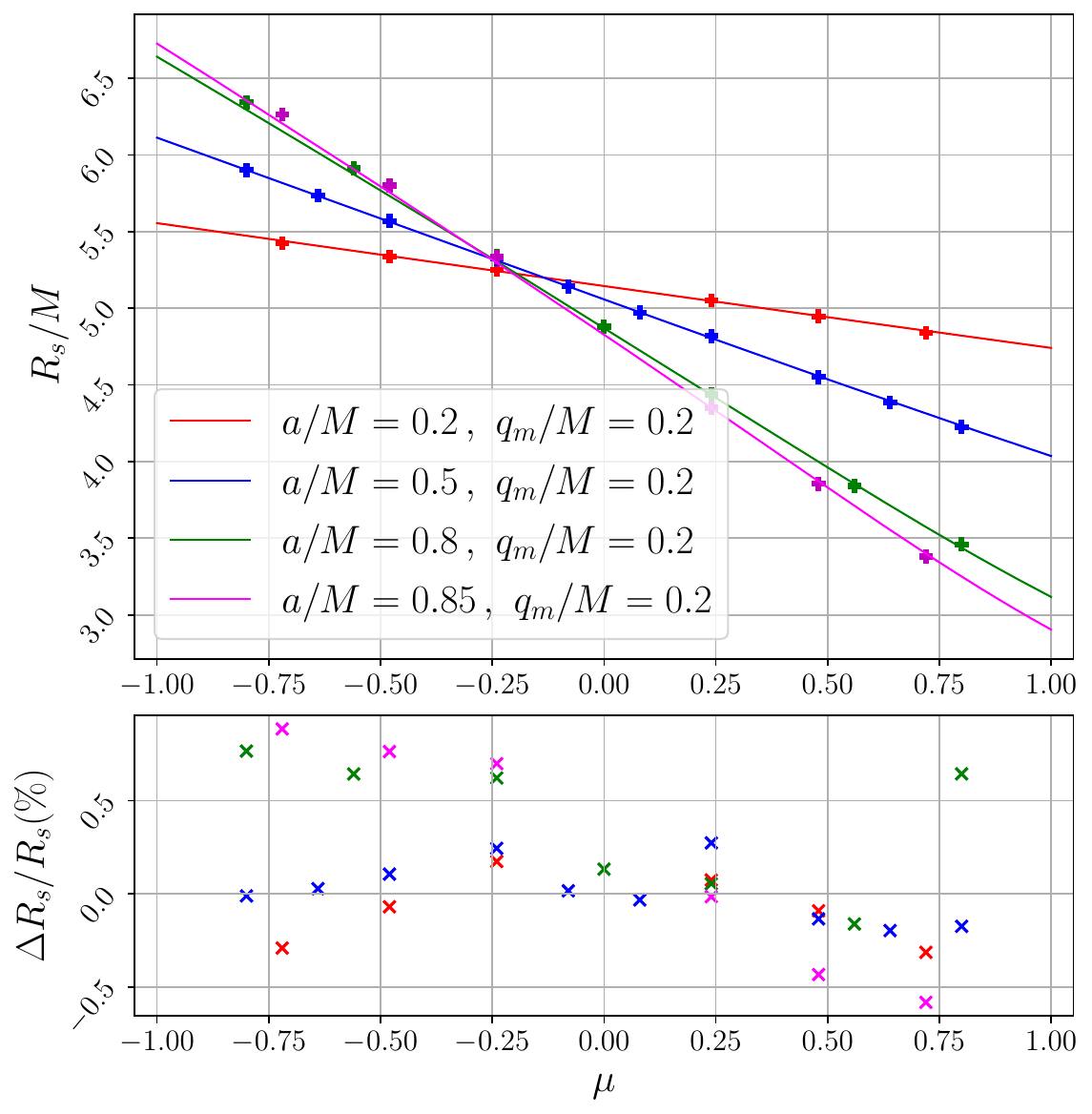}
\caption{Explicit tests of the eikonal quasinormal modes-shadow correspondence for rotating regular Bardeen black holes, for different values of the regularization parameter $q_m$ at fixed angular momentum $a$ ($a/M=0.2$, left panels), and likewise for different values of the angular momentum at fixed regularization parameter $q_m$ ($q_m/M=0.2$, right panels). In both cases, solid curves indicate the BH shadow radius $R_s(\mu)$ computed using closed photon orbits [Eq.~(\ref{eq:rs})] after making use of the the Bohr-Sommerfeld quantization condition [Eq.~(\ref{eq:BScondition})], whereas the points (diamond markers for fixed spin, plus sign markers for fixed hair parameter) are obtained from eikonal QNMs making use of the QNM-shadow correspondence [Eq.~(\ref{eq:correspondencenew})]. The lower panels display percentage residuals between the two different calculations: the difference between the two is always within the $\lesssim 1\%$, hence within the numerical precision of the QNM code.}
\label{fig:bardeen}
\end{figure*}

Our goal is now to generalize the procedure outlined in Sec.~\ref{sec:kerr} for the Kerr metric and based on the work of Ref.~\cite{Yang:2021zqy} to a wider class of rotating (regular) black holes, specializing to the rotating Bardeen and Hayward BHs discussed in Sec.~\ref{sec:regular}. We recall that the connection is established by combining Eqs.~(\ref{eq:expansionnew},\ref{eq:connection}): the former connects the shadow radius to parameters characterizing photon orbits, while the latter makes the connection to eikonal quasinormal modes explicit. We now dig deeper into the steps leading to this connection, to better understand their regime of validity.

Let us turn our attention to Eq.~(\ref{eq:connection}), which is valid in general for rotating BHs whose Hamilton-Jacobi equation of motion is separable. On the other hand, Eq.~(\ref{eq:expansion}) would only seem to apply to Kerr BHs. In fact, it has been derived from the Bohr-Sommerfeld quantization condition given by Eq.~(\ref{eq:BScondition}), which in turn was obtained via a WKB-like analysis of the angular part of the Teukolsky equation for the Kerr metric~\cite{Yang:2012he}. However, the crucial point is to notice that the angular part of the Teukolsky equation~\cite{Teukolsky:1972my} reduces, in the eikonal limit $(\ell \gg 1)$, to the angular part of the Klein-Gordon equation for a massless scalar field $\chi$:
\begin{eqnarray}
&&F_{_{\ell m}}\chi_{\theta} = 0\,, \nonumber \\
&&F_{_{\ell m}} \equiv \frac{1}{\sin\theta}\frac{d}{d\theta} \left ( \sin\theta \frac{d}{d\theta}\right ) \nonumber \\
&&+ \left [ a^2\omega^2\cos^2\theta - \frac{m^2}{\sin^2\theta} - A_{\ell m} \right ] \,.
\label{eq:KGang-part}
\end{eqnarray}
This implies that the angular part of the perturbation equation for Kerr BHs is universal, i.e.\ independent of the spin of the perturbation.

Driven by the above considerations, it appears reasonable to push these ideas further, leading us to expect that the procedure described by Sec.~\ref{sec:kerr} is applicable \textit{to any rotating metric for which both the Hamilton-Jacobi and Klein-Gordon equations are separable assuming that the following asymptotic behaviour holds}:
\vskip 0.15cm
\framebox[\linewidth]{Teukolsky-like equation $\underset{\ell \gg 1}{\longrightarrow}$ Klein-Gordon equation}
\vskip 0.15cm
\noindent The idea is then to use the angular part of the Klein-Gordon equation to derive an analog of the B-S quantization condition Eq.~(\ref{eq:BScondition}) for the metric under consideration, from which equations analogous to Eqs.~(\ref{eq:expansion},\ref{eq:correspondence}) can be directly derived. Under the assumptions described above, the very same steps required to establish the QNM-shadow correspondence for Kerr BHs~\cite{Yang:2021zqy} can therefore be transposed to the axisymmetric metric which is being considered.

The natural question to ask is then: which metrics satisfy the above separability conditions? To answer this question, we once more return to the Newman-Janis algorithm, and in particular to the important work of Chen \& Chen~\cite{Chen:2019jbs}: this work showed that the most general stationary, axisymmetric BH metric which can be obtained applying the Newman-Janis algorithm (without complexification~\cite{Azreg-Ainou:2014aqa,Azreg-Ainou:2014pra}) is described, in Boyer-Lindquist coordinates, by the following line element~\cite{Chen:2019jbs}:
\begin{eqnarray}
ds^2 = &-&Bdt^2 + 2a\sin^2\theta \left ( B - \sqrt{\frac{B}{A}} \right ) dtd\phi \nonumber \\
&+& \Psi d\theta^2 + \frac{\Psi}{W}dr^2 \nonumber \\
&+& \sin^2\theta \left [ \Psi + a^2\sin^2\theta \left ( 2\sqrt{\frac{B}{A}} - B \right ) \right ] d\phi^2\,, \nonumber \\
\label{eq:general}
\end{eqnarray}
where $A(r,\theta)$, $B(r,\theta)$, $\Psi(r,\theta)$, and $W(r,\theta)$ are generic functions of $r$ and $\theta$, subject to the constraint:
\begin{eqnarray}
W(r) \equiv A(r,\theta)\Psi(r,\theta) + a^2\sin^2\theta \,.
\label{eq:wconstraint}
\end{eqnarray}
The space-time metric described by Eq.~(\ref{eq:general}), with the constraint given by Eq.~(\ref{eq:wconstraint}), is very general, and reduces to the Kerr space-time upon identification of $B(r,\theta)=A(r,\theta)= 1-2Mr/\Psi$, with $\Psi = r^2 + a^2\cos^2\theta$. In the same work, the conditions required in order for the space-time described by Eq.~(\ref{eq:general}) to admit separable Hamilton-Jacobi and Klein-Gordon equations are also discussed, and are respectively given by~\cite{Chen:2019jbs}:
\begin{eqnarray}
\Psi(r,\theta) &=& \zeta(r) + \varpi(\theta)\,, \label{eq:cond1} \\
Y &\equiv& \sqrt{\frac{B(r,\theta)}{A(r,\theta)}} = Y(r)\,, \quad Y(\infty) \to 1\,,
\label{eq:cond2}
\end{eqnarray}
where $\zeta(r)$ and $\varpi(\theta)$ are generic functions of $r$ and $\theta$ respectively. The requirement that $\Psi(r,\theta)$ is additively separable [Eq.~(\ref{eq:cond1})] ensures the separability of the Hamilton-Jacobi equation (but is not, in general, required to ensure the separability of the null geodesic equation), whereas the second condition [Eq.~(\ref{eq:cond2})] ensures the separability of the Klein-Gordon equation. Moreover, Ref.~\cite{Chen:2019jbs} explicitly provides the angular part of the Klein-Gordon equation for the generic metric given by Eq.~(\ref{eq:general}), and it is straightforward to show that it takes the same form as Eq.~(\ref{eq:KGang-part}). This implies that we can derive an analog of the B-S quantization condition [Eq.~(\ref{eq:BScondition})] for BHs described by the metric in Eq.~(\ref{eq:general}), thereby generalizing the correspondence described in Sec.~\ref{sec:correspondence} for the Kerr metric.

\begin{figure*}[!ht]
\includegraphics[width=0.49\linewidth]{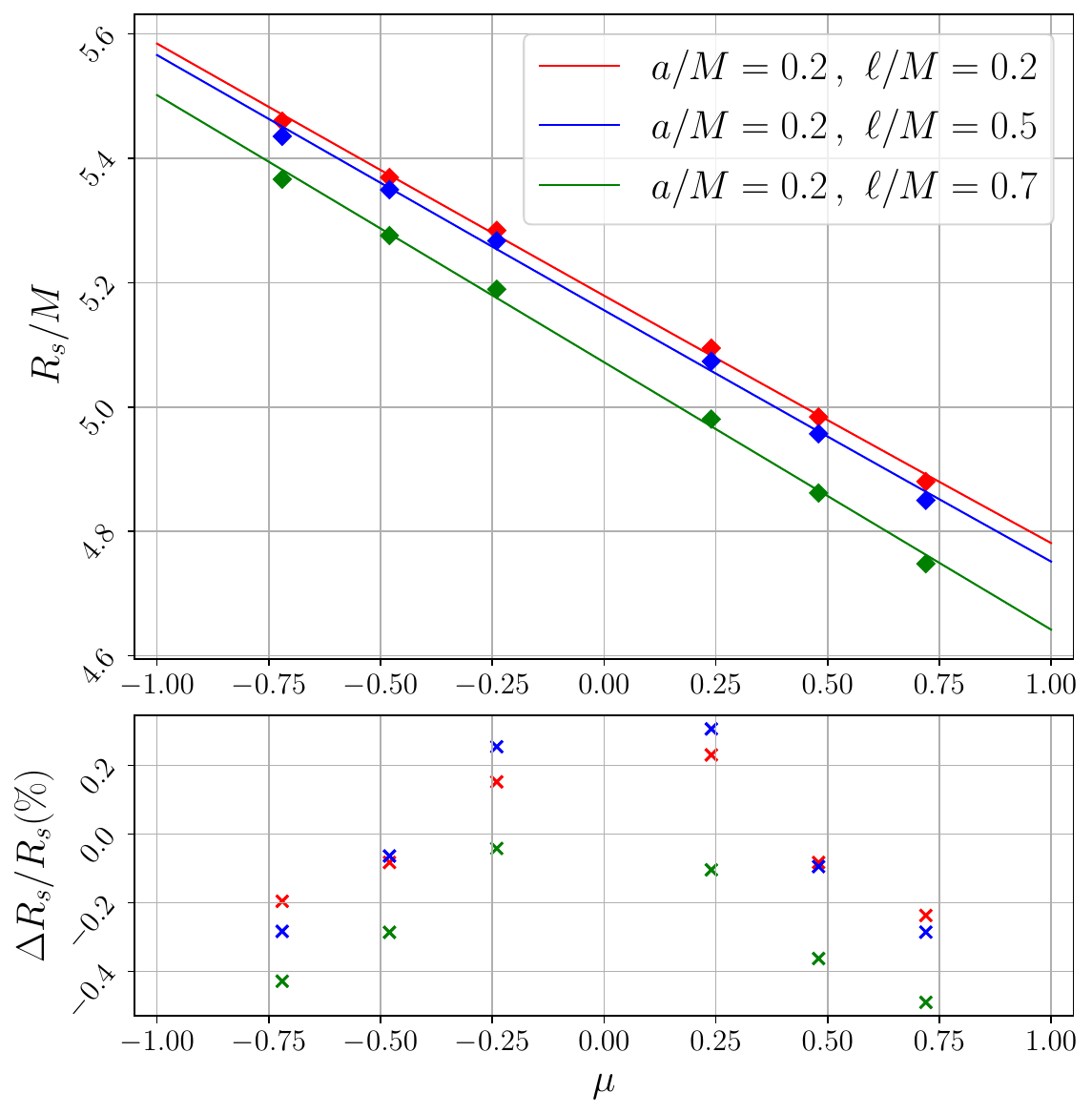}\,\,\includegraphics[width=0.49\linewidth]{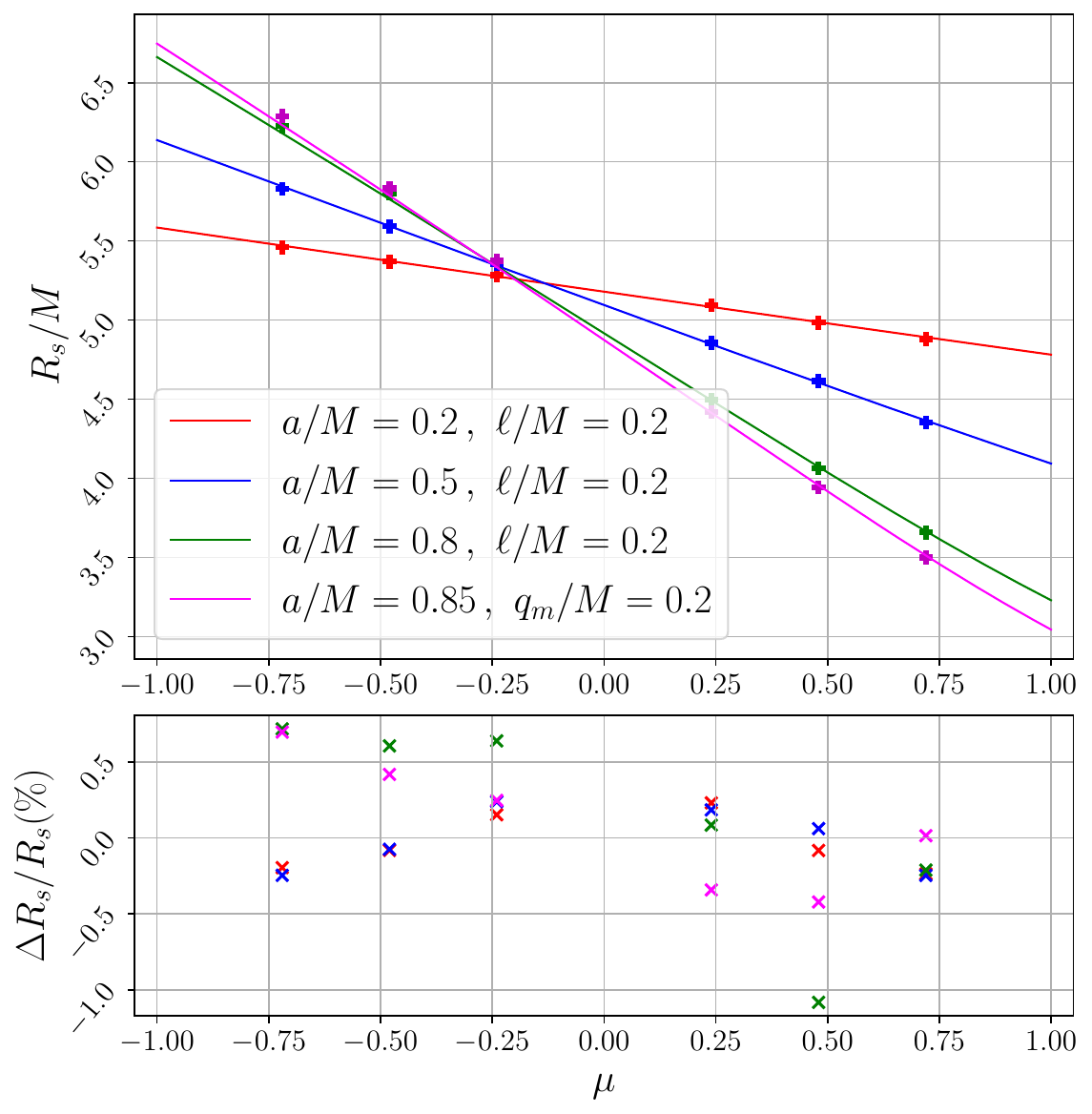}
\caption{As in Fig.~\ref{fig:bardeen} but focusing on the rotating regular Hayward BH, and the associated regularization parameter $\ell$.}
\label{fig:hayward}
\end{figure*}

To summarize, the above physical and mathematical considerations lead us to conclude that the eikonal QNM-shadow radius correspondence identified in Ref.~\cite{Yang:2021zqy} and described in Sec.~\ref{sec:kerr} can be extended to space-time metrics described by Eq.~(\ref{eq:general}), subject to the constraints given by Eqs.~(\ref{eq:cond1},\ref{eq:cond2}). It is straightforward to check that the rotating Bardeen and Hayward space-times, with line elements described by Eqs.~(\ref{eq:bardeen},\ref{eq:bardeenfunction},\ref{eq:haywardfunction}), satisfy these conditions.

To further cement the above statement, we explicitly verify that the eikonal QNM-shadow radius correspondence outlined earlier holds for the rotating Bardeen and Hayward BHs. On the one hand, we calculate eikonal QNMs for these space-times adopting $2+1$ time-evolution code, based on the Prony method~\cite{Berti:2007dg} and briefly described in Appendix~\ref{sec:appendixa}, with the explicit form of the wave equation provided in Appendix~\ref{sec:appendixb}. On the other hand, the shadow radii are calculated by explicitly solving for closed photon orbits, with a procedure briefly described in Appendix~\ref{sec:appendixc}. This procedure allows us to obtain $R_s(\mu)$, as we are able to associate each photon orbit to a specific value of $\mu$. The eikonal QNMs are then themselves used to obtain a value of $R_s(\mu)$ through Eq.~(\ref{eq:correspondencenew}). The two values of $R_s(\mu)$, from eikonal QNMs and closed photon orbits, are then compared to numerically gauge the accuracy of the correspondence.

The results of this comparison are summarized in Fig.~\ref{fig:bardeen} and Fig.~\ref{fig:hayward}, for the rotating Bardeen and Hayward BHs respectively. In particular, for each figure the upper left panel (with diamond markers) focuses on the effect of increasing the ``hair parameter'' ($q_m/M\,,\ell/M=0.2,0.5,0.7$ for both the Bardeen and Hayward BHs in red, blue, and green respectively) at fixed value of the BH spin ($a/M=0.2$), whereas the upper right panel (with plus sign markers) focuses on the effect of increasing the value of the BH spin ($a/M=0.2,0.5,0.8,0.85$) at fixed value of the hair parameter ($q_m,\,\ell=0.2M$). In each panel, solid curves represent $R_s(\mu)$ as computed using closed photon orbits, in particular by solving the B-S quantization condition Eq.~(\ref{eq:BScondition}) and then using Eq.~(\ref{eq:rs}), see Appendix~\ref{sec:appendixc} for details. On the other hands, (discrete) dots are obtained computing eikonal QNMs and translating them to values of $R_s(\mu)$ using Eq.~(\ref{eq:correspondencenew}). The lower right panels instead show the (percentage) relative deviation in $R_s(\mu)$ as calculated using the two methods. We recall that $\mu=0$ corresponds to the two points on the celestial $y$ axis~\cite{Perlick:2021aok}.

\textbf{The points we sampled to test the eikonal QNM-shadow correspondence for rotating regular BHs are also shown in Fig.~\ref{fig:bardeen_hayward_bh_nobh}, with the same choice of colors and markers as in Figs.~\ref{fig:bardeen} and~\ref{fig:hayward}. It is worth noting from Fig.~\ref{fig:bardeen_hayward_bh_nobh} that the concept of extremality for these rotating BHs is inherently two-dimensional. One should therefore look at spin and hair parameter simultaneously, and not in isolation, in order to gauge how close to extremality a given BH is. In fact, the point with $a/M=0.2$ and $\{q_m/M,\ell/M\}=0.7$, as well as the point with $a/M=0.85$ and $\{q_m/M,\ell/M\}=0.2$, are both very close to being extremal despite their low values of spin and hair parameter respectively. Obviously in practice we can only test the eikonal QNM-shadow correspondence for a discrete selection of points in parameter space, but Fig.~\ref{fig:bardeen_hayward_bh_nobh} makes it clear that our choice also includes near-extremal rotating regular BHs.}

As is visually clear, Fig.~\ref{fig:bardeen} and Fig.~\ref{fig:hayward} show \textit{excellent} agreement between $R_s(\mu)$ as computed using eikonal QNMs and closed photon orbits. From the lower panel we indeed see that the relative deviation between the two is essentially always below the $\%$ level, \textbf{even for near-extremal rotating Bardeen and Hayward BHs (see the green diamond marker and green plus sign marker in Fig.~\ref{fig:bardeen_hayward_bh_nobh})}. The deviations we find are largest for the rotating Hayward BH at high spin values, whereas for low spin values the relative deviations for both the rotating Bardeen and Hayward BHs are always $\lesssim 0.4\%$ in absolute value. These numbers can be considered indicators of an excellent agreement between the two routes towards calculating $R_s(\mu)$, given the different sources of numerical error potential at play. Firstly, the numerical precision of the code we used to implement the Prony method is of order $\sim 1\%$. This precision figure has been ``calibrated'' to the Kerr metric, by comparing the QNM frequencies obtained using our method against the state-of-the-art reported by Emanuele Berti.~\footnote{See~\url{https://pages.jh.edu/eberti2/ringdown/}.} We find a very good agreement with Berti's QNMs, with relative deviations of $\lesssim 0.2\%$ for slowly rotating BHs $(a/M \lesssim 0.5)$, whereas the relative deviations increase to to $\lesssim 1-1.5 \%$ for values of $a/M$ approaching extremality: this is due to the mode coupling phenomenon~\cite{Burko:2013bra}, which complicates the process of QNM extraction. Although this is by no means a formal proof, the above considerations, coupled to the Kerr-like form of the rotating Bardeen and Hayward space-times [Eq.~(\ref{eq:rotatingbh})], lead us to expect that the numerical uncertainty floor of our calculated QNMs is of the $\%$ level. Earlier studies on $2+1$ time-evolution codes indeed suggests that their precision is of order $\%$ or better~\cite{Doneva:2020nbb,Doneva:2020kfv}. Therefore, the agreement between the two methods for computing $R_s(\mu)$ as displayed in Fig.~\ref{fig:bardeen} and Fig.~\ref{fig:hayward} being always within $\lesssim 1\%$ can be considered excellent, as it always lies within the numerical uncertainty floor of our method. Moreover, although establishing the correspondence requires us to calculate eikonal QNMs, numerically speaking we cannot of course take $\ell \to \infty$. For our purposes, we have verified that at $\ell=12$ our QNMs have basically ``converged'' to the eikonal limit within the precision of the code, with no appreciable improvements observed when taking larger values of $\ell$, although of course our choice of finite (but relatively large) value of $\ell=12$ introduces a further source of numerical uncertainty.

\textbf{Before closing, we wish to briefly discuss a few caveats and potential limitations of our treatment. Firstly, while we have remained agnostic as to the physical origin of the Bardeen and Hayward BHs, it is known that both solutions can emerge within theories of non-linear electrodynamics. In such theories, photons move along null geodesics of an \textit{effective} metrics which accounts for the modified electrodynamics sector, and this alters the shadow computation relative to the standard one, potentially invalidating the QNM-shadow correspondence. Here we have chosen to follow the same phenomenological approach adopted in Ref.~\cite{Vagnozzi:2022moj}, remaining agnostic as to the physical origin of these metrics, given \textit{a)} the many different potential origins thereof (not all of which are rooted within non-linear electrodynamics), and \textit{b)} the difficulty in constructing a \textit{well-motivated, first-principles} non-linear electrodynamics Lagrangian which gives rise to these metrics (the theory is usually reverse-engineered, see Ref.~\cite{Bokulic:2023afx}). We refer the reader to Sec.~CB of Ref.~\cite{Vagnozzi:2022moj} for further discussions on this rationale.}

\textbf{Another potential limitation of our treatment lies in the use of the Newman-Janis algorithm in obtaining the rotating versions of the regular BHs we studied. While widely used, the algorithm has received criticisms due to ambiguities in the complexification procedure, which in our case are addressed by adopting the complexification-free version of Refs.~\cite{Azreg-Ainou:2014aqa,Azreg-Ainou:2014pra}. Another more serious concern lies in the fact that the algorithm works well for vacuum solutions, but does not necessarily work for metrics sourced by external fields (including the effective source generated by non-linear electrodynamics), see e.g.\ Refs.~\cite{Drake:1998gf,Rajan:2015ffs,Rajan:2016zmq,Arkani-Hamed:2019ymq,Guevara:2020xjx,Mazza:2021rgq,Kamenshchik:2023woo} for discussions on some of its limitations. Nevertheless, our phenomenological approach, where we remain agnostic as to the origin of the Bardeen and Hayward BHs (see also the discussion in the paragraph above), make these limitations less worrisome. In this perspective, the Newman-Janis algorithm has actually proven to be an extremely useful tool, by providing a very general class of axisymmetric metrics, irrespective of their origin, which have helped us identify a set of conditions under which the eikonal QNM-shadow correspondence holds.}

To wrap up the discussion, we find excellent agreement between $R_s(\mu)$ as computed using eikonal QNMs and closed photon orbits, relating the two via Eq.~(\ref{eq:correspondencenew}). These results further cement the validity of the eikonal QNM-shadow radius correspondence~\cite{Yang:2021zqy} for the two rotating BHs discussed. Nevertheless, the mathematical and physical arguments presented earlier lead us to expect that such a correspondence should hold \textit{at the very least} for a wide class of axisymmetric metrics (regular and non), going beyond the two representative and interesting examples we considered here.

\section{Conclusions}
\label{sec:conclusions}

We are now living in an exciting era where black holes have moved from being mere mathematical objects, to having their effects being routinely observed in a wide range of channels. Motivated by this, and the increasing attention towards observational tests of gravity and fundamental physics in the strong-field regime, we have further explored the connection between two BH-related observables: eikonal ($\ell \gg 1$) quasinormal modes and shadow radii. More specifically, we studied in detail the correspondence between the real part of eikonal QNMs and BH shadow radii in the case of rotating (axisymmetric) space-times, which was only partially dealt with in earlier studies. In particular, Jusufi~\cite{Jusufi:2020dhz} only studied the case of equatorial ($m=\pm\ell$) QNMs, therefore with a restrictive definition of shadow radius, whereas Yang~\cite{Yang:2021zqy} identified a more general correspondence, but limited to the Kerr metric. We have argued, based on mathematical and physical arguments, that the correspondence identified by Yang in Ref.~\cite{Yang:2021zqy} [see Eq.~(\ref{eq:correspondencenew})] should apply to a wide range of axisymmetric metrics, subject to requirements on the separability of the Hamilton-Jacobi equation for null geodesics and the Klein-Gordon equation. From the practical point of view, axisymmetric metrics obtained from the Newman-Janis algorithm fall within the space-times to which the correspondence applies, provided the requirements set out in Eqs.~(\ref{eq:wconstraint}--\ref{eq:cond2}) are applied [with the metric expressed as in Eq.~(\ref{eq:general})]. We have then explicitly verified that the correspondence holds for two well-studied regular BH space-times: the rotating Bardeen and Hayward BHs. \textbf{We stress that our treatment of rotating regular BHs may have some limitations, due to both the potential physical origin of the Bardeen and Hayward BHs (which we have remained agnostic about, but could be rooted within theories of non-linear electrodynamics), as well as limitations to the validity of the Newman-Janis algorithm, which nonetheless has proven to be an extremely useful tool in generating a general class of axisymmetric metrics. We redirect the reader to the end of Sec.~\ref{sec:correspondence} for further discussions on these limitations.}

Circling back to the title of our work, we have therefore confirmed that there is a direct relation between what one hears from BHs (gravitational waves, whose ringdown part is directly connected to QNMs) and what one sees of BHs. The natural question is, of course, whether such a correspondence is amenable to observational tests. A simple answer is: at some point, but probably not in the near future. As argued in Ref.~\cite{Yang:2021zqy}, the significant challenge here is to identify a system which allows for simultaneous horizon-scale VLBI interferometry \textit{and} GW ringdown measurements, as systems which are favorable for one tend to not be for the other. For instance, the extreme mass-ratio inspirals which LISA will observe should be accompanied by accretion disks which would facilitate VLBI observations~\cite{Pan:2021ksp} -- however, their faint ringdown signals do not make them ideal targets for GW spectroscopy~\cite{Baibhav:2019rsa}. On the other hand, LISA is also expected to detect several (loud) massive BH mergers~\cite{Klein:2015hvg,Bonetti:2018tpf}, which are ideal targets for GW spectroscopy and could possibly be embedded in a gas-rich environment -- however, these sources would also be at cosmological distances, with angular sizes at best of order ${\cal O}({\text{nas}})$, a few orders of magnitude below the current sensitivity. Future improvements both on the VLBI side (either expanding EHT's terrestrial footprint~\cite{Ayzenberg:2023hfw}, upgrading with additional telescopes from space~\cite{Haworth:2019urs}, moving to the optical band~\cite{Tsupko:2019pzg}, or even more futuristically using constellations of satellites to perform X-ray interferometry~\cite{Uttley:2019ngm}) and on the GW side (for instance through deep next-generation space-based GW detectors such as the recently proposed AMIGO~\cite{Baibhav:2019rsa,Ni:2019nau}) could help with the quest, although a complete forecast would require a dedicated study which goes beyond the scope of the present work.

There are many interesting follow-up directions one could pursue starting from our work. Firstly, it would be worth understanding if a more mathematical-grounded justification of our guess in Sec. \ref{sec:correspondence}, regarding the eikonal behavior of the Teukolsky equation, can be found. It would also be interesting to investigate whether the separability conditions we have outlined are necessary or merely sufficient, or in other words, if the correspondence identified applies to a wider class of axisymmetric BHs. A rather easy exercise would instead be to explicitly test the correspondence on other metrics of interest, e.g.\ among the ones whose shadow properties were studied in Ref.~\cite{Vagnozzi:2022moj}: some of these can be expected to potentially violate the eikonal correspondence, due to non-minimal couplings of photons to other fields. Another interesting extension could instead consider photon ring observables~\cite{Johnson:2019ljv}, and thereby the correspondence between the imaginary part of eikonal QNMs, and the relative brightness of photon rings one can observe in VLBI images. Of course, the correspondence opens up to novel tests of GR and fundamental physics, along the lines discussed in Ref.~\cite{Chen:2022nlw}, and it would be interesting to further study these, especially for cases where the eikonal correspondence is potentially violated (see e.g.\ Refs.~\cite{Konoplya:2017wot,Glampedakis:2019dqh,Chen:2019dip,Silva:2019scu,Chen:2021cts,Li:2021mnx,Moura:2021eln,Bryant:2021xdh,Nomura:2021efi,Guo:2021enm,Konoplya:2022gjp}), or examining how well the connection holds for moderate $\ell$ (see e.g.\ Ref.~\cite{Volkel:2020xlc}). Moreover, all the cases we have studied featured a single-peak effective potential, but it could be interesting to extend the study the case of potentials with multiple peaks (see e.g.\ explicit hairy BH solutions in Refs.~\cite{Gan:2021pwu,Gan:2021xdl,Guo:2022umh}), as well as horizonless objects: in the former case, the existence of long-lived and sub-long-lived modes are expected to give rise to echo phenomena, which might themselves require a partial rethinking of the QNM-shadow correspondence. Additionally, a concrete forecast for the possibility of a direct multi-messenger test of the QNM-shadow correspondence from (far-)future GW spectroscopy and VLBI imaging, as anticipated earlier, would certainly be very interesting. Finally, a similar correspondence between QNMs and greybody factors has recently been pointed out~\cite{Konoplya:2024lir,Konoplya:2024vuj}, suggesting a connection between shadows and QNM-related observables, and features related to BH evaporation. This connection deserves future study. We leave the study of these and related interesting questions to future work.

\begin{acknowledgments}
\noindent We thank Kyriakos Destounis, Kostas Kokkotas, and Christian Kr\"{u}ger for many very useful discussions throughout the project. D.P. and S.V. acknowledge support from the Istituto Nazionale di Fisica Nucleare (INFN) through the Commissione Scientifica Nazionale 4 (CSN4) Iniziativa Specifica ``Quantum Fields in Gravity, Cosmology and Black Holes'' (FLAG). S.V. acknowledges support from the University of Trento and the Provincia Autonoma di Trento (PAT, Autonomous Province of Trento) through the UniTrento Internal Call for Research 2023 grant ``Searching for Dark Energy off the beaten track'' (DARKTRACK, grant agreement no.\ E63C22000500003). This publication is based upon work from the COST Action CA21136 ``Addressing observational tensions in cosmology with systematics and fundamental physics'' (CosmoVerse), supported by COST (European Cooperation in Science and Technology).
\end{acknowledgments}

\appendix

\section{QNM calculation}
\label{sec:appendixa}

To obtain the eikonal QNMs for the rotating Bardeen and Hayward BHs, which we then translate to $R_s(\mu)$ through the eikonal QNM-shadow correspondence (dots in the upper panels of Fig.~\ref{fig:bardeen} and Fig.~\ref{fig:hayward}), we adopt a time-evolution method, by numerically integrating the Klein-Gordon equation, which we assume to be, according to our guess in Sec. \ref{sec:correspondence}, the eikonal limit of the Teukolsky-like equation for the field $\Phi$, in a spacetime described by the metric \eqref{eq:general}:
\begin{eqnarray}
\Box \Phi = \nabla^{\mu}\nabla_{\mu} \Phi = \frac{1}{\sqrt{-g}}\left[\sqrt{-g} g^{\mu\nu}\Phi_{,\nu}\right]_{,\mu} = 0\,,
\label{eq:KG}
\end{eqnarray}
to which we impose Sommerfeld boundary conditions. We solve Eq.~(\ref{eq:KG}) using the method of lines and a fourth order Runge-Kutta time integrator. The axisymmetric nature of the space-times we consider allows for separation of variables, assuming the following ansatz:
\begin{eqnarray}
\Phi(t,r,\theta,\phi) = e^{im\phi}\Psi(t,r,\theta)\,,
\label{eq:phiansatz}
\end{eqnarray}
which cleanly separates out the $\phi$-dependence of the scalar field.~\footnote{To be precise, for stability reasons the evolution is carried out in $(t,r_{\star},\theta,\phi_*)$ coordinates~\cite{Krivan:1996da}, where $d\phi_{\star} = d\phi + dr (a/\Delta)$ and $dr_{\star} = dr(r^2 + a^2)/\Delta$. Therefore, the actual ansatz for separation of variables is $\Phi(t,r_{\star},\theta,\phi_*) = e^{im\phi_{\star}}\Psi(t,r_{\star},\theta)$.} This ansatz turns the problem into a 2+1 time-evolution problem.

We evolve $\Psi$ over a two-dimensional $(r_{\star},\theta)$ grid, where $r_{\star}$ is the so-called tortoise coordinate (see footnote~15), in principle covering the whole range $r_{\star} \in (-\infty, +\infty)$, while the angular coordinate extends over the half plane $\theta \in [0,\pi)$. Absorbing boundary conditions at the event horizon and at spatial infinity are applied following Ref.~\cite{Ruoff:2000nj}, while continuity conditions at $\theta = \pi$ are imposed such that $\partial_{\theta}\Psi(\theta \in \{0,\pi\}) = 0$ for $m=0$, and $\Psi(\theta \in \{0,\pi\}) = 0$ for $m\neq 0$, following Ref.~\cite{Dolan:2011dx}. In practice, to avoid spurious reflections from the boundaries due to the finite numerical precision, the numerical values of $r_{\star \pm\infty} = \pm \infty$ are chosen in such a way that they are as far away as possible from the observer $r_{\star O}$: in practice, we have verified that $r_{\star O} \sim 30M$ is accurate enough for the purposes of recording a sufficiently long signal, free from unwanted reflections. The signal in time domain is then processed via the Prony method to extract the desired QNM frequencies~\cite{Berti:2007dg}.

Following Refs.~\cite{Doneva:2020nbb,Zhang:2020pko,Zhang:2021btn}, we set the initial conditions for the scalar field as follows:
\begin{eqnarray}
\Psi(t=0) \sim Y_{\ell m}e^{-\frac{(r_{\star O} - r_\star)^2}{2\sigma^2}}\,,
\label{eq:initialconditions}
\end{eqnarray}
where $\sigma$, which we set to $\sigma = 1M$, is the width of the initial (Gaussian wave) displacement, and $Y_{\ell m}$ is the $\theta$-dependent part of the $(\ell,m)$ spherical harmonic. Even though the Klein-Gordon equation for Kerr BHs, as well as for the RBHs under consideration, does not show an explicit dependence on $\ell$, the initial conditions given in Eq.~(\ref{eq:initialconditions}) are found to predominantly excite the $(\ell,m)$ mode~\cite{Kokkotas:1999bd}. This has been explicitly checked for the Kerr space-time, by comparing our QNM frequencies against the state-of-the-art reported by Emanuele Berti on~\url{https://pages.jh.edu/eberti2/ringdown/}, leading to the numerical precision figures quoted in Sec.~\ref{sec:correspondence}. Finally, the explicit form of the Klein-Gordon equation [Eq.~(\ref{eq:KG})] for the rotating regular space-times considered was computed using a symbolic calculation routine implemented in \texttt{Maple}, and is provided in Appendix~\ref{sec:appendixb}.

\section{Klein-Gordon equation}
\label{sec:appendixb}

Below we report the explicit form of the Klein-Gordon equation [Eq.~\ref{eq:KG}] describing the evolution of a perturbation evolving on the background of the rotating Bardeen and Hayward space-times. For a general axisymmetric metric taking the form given by Eq.~(\ref{eq:rotatingbh}), the Klein-Gordon equation can be written as follows:
\begin{widetext}
\begin{equation}
\begin{split}
\Box \Phi &= -\frac{\sigma\Sigma^2}{A} \frac{\partial^2 \Phi}{\partial t^2} + \frac{(r^2 + a^2)^2}{\Delta}\frac{\partial^2 \Phi}{\partial r_*^2} + \left[  -\frac{B(a^2 + r^2)\partial_r\Sigma}{2A\Sigma} \right. - \frac{(a^2 + r^2)\partial_r\Delta}{2\Delta} + 2r\\
& \left. + \frac{(a^2 + r^2)(\Sigma - 2f)\partial_r\Sigma}{2A} - \frac{(a^2 + r^2)C\partial_rf}{A} \right]\frac{\partial \Phi}{\partial r_*} - \frac{4af\Sigma^2}{A} \frac{\partial^2 \Phi}{\partial t\partial \phi_*} + \frac{\partial^2 \Phi}{\partial \theta^2}\\
&+ \left[-\frac{B\partial_{\theta}\Sigma}{2A\Sigma} + \frac{(\Sigma - 2f)\partial_{\theta}\sigma}{2A} + \frac{D\cos\theta}{A\sin\theta}\right]\frac{\partial \Phi}{\partial \theta} + \left[ - \frac{Ba\partial_r\Sigma}{2A\Sigma} - \frac{a\partial_r\Delta}{2\Delta} \right.\\
&+ \left. \frac{a(\Sigma - 2f)\partial_r\sigma}{2A} - \frac{aC\partial_r f}{A} \right] \frac{\partial \Phi}{\partial \phi_*} + \frac{E}{\sin^2\theta A \Delta}\frac{\partial^2 \Phi}{\partial \phi_*^2} + \frac{2a(a^2 + r^2)}{\Delta}\frac{\partial^2 \Phi}{\partial \phi_* \partial r_*} = 0\,,
\end{split}
\label{appF:WE1}
\end{equation}
where we have defined:
\begin{align}
&\Sigma(r,\theta) \equiv r^2 + a^2\cos^2\theta\,,\\
&\sigma(r,\theta) \equiv (r^2 + a^2)^2 -a^2\Delta\sin^2\theta\,,\\
&A(r,\theta) \equiv 4a^2f^2\sin^2\theta + \sigma(\Sigma - 2f)\,,\\
&B(r,\theta) \equiv 8a^2f^2\sin^2\theta + \sigma(\Sigma -4f)\,,\\
&C(r,\theta) \equiv -4fa^2\sin^2\theta\,,\\
&D(r,\theta) \equiv 8 f^2a^2\sin^2\theta + \sigma(\Sigma - 2f)\,,\\
&E(r,\theta) \equiv 4f^2a^4(1 - \cos^2\theta\cos2\theta f^2a^4) - (\sin^2\theta\sigma^2 a^2 + \Delta\Sigma^2)(2f - \Sigma)\,,
\end{align}
\end{widetext}
and the differences between the rotating Bardeen and Hayward space-times are all encapsulated in the function $f(r)$, see the discussion in Sec.~\ref{sec:regular}, and in particular the metric given by Eq.~(\ref{eq:rotatingbh}), as well as the explicit forms of $F(r) \equiv 1-2f(r)/r^2$ given in Eqs.~(\ref{eq:fb},\ref{eq:fh}).

\section{Shadow radius calculation from photon orbits}
\label{sec:appendixc}

The shadow radius $R_s(\mu)$ given by the solid curves in Fig.~\ref{fig:bardeen} and Fig.~\ref{fig:hayward} is computed through the canonical approach, i.e.\ studying closed photon orbits. The difference with respect to the usual approach is that we also need to compute the value of $\mu$ associated to each closed photon orbit, and thereby each point on the shadow, in order to establish the connection to eikonal QNMs. The code we developed numerically solves the B-S quantization condition Eq.~(\ref{eq:BScondition}) for every set of conserved quantities $(E,{\cal C},L_z)$ corresponding to closed photon orbits, in order to determine $L$ and therefore $\mu = L_z/L$. As per Eq.~(\ref{eq:rs}), the shadow radius is then given by $R_s = \sqrt{{\cal C} + L_z^2}/E$ which, since $L$ is now known, puts us in the position to directly determine $R_s(\mu)$.

We recall that for Kerr BHs $L_z$ and ${\cal C}$ are related to the parameter of closed photon orbits $r$ via~\cite{Yang:2012he}:
\begin{align}
&\eta \equiv \frac{{\cal C}}{E^2} = -\frac{r^3(r^3 - 6Mr^2 + 9M^2r - 4a^2M)}{a^2(r-M)^2}\,, \\
&\xi \equiv \frac{L_z}{E} = - \frac{r^3 - 3Mr^2 + a^2r + a^2M}{a(r-M)}\,.
\label{eq:clzkerr}
\end{align}
However, this is no longer true for the rotating Bardeen and Hayward BHs, where the corresponding expressions are found by solving:
\begin{eqnarray}
\mathcal{R}(r) = 0\,, \qquad \frac{d\mathcal{R}(r)}{dr} = 0\,, \qquad \frac{d^2\mathcal{R}(r)}{dr} > 0\,,
\label{eq:rrdrr}
\end{eqnarray}
where $\mathcal{R}(r)$ is the function such that the radial geodesic equation can be written as:
\begin{eqnarray}
\frac{dr}{d\tau} = \pm \frac{E}{\Sigma}\sqrt{\mathcal{R}(r)}\,,
\label{eq:drdtau}
\end{eqnarray}
with $\tau$ being the affine parameter along the geodesic. For the rotating Bardeen and Hayward BHs, the result of this calculation can be found in Refs.~\cite{Tsukamoto:2014tja,Abdujabbarov:2016hnw} (see also Ref.~\cite{Tsukamoto:2017fxq}), whose results we quote here:
\begin{widetext}
\begin{eqnarray}
\eta &=& -\frac{r^3 \left [ (1+m'^{2})r^3+2m'r(r^2-3mr+2a^2)-m(6r^2-9mr+4a^2) \right ] }{a^2 \left [ m+r(m'-1) \right ] ^2}\,, \label{eq:eregular} \\
\xi &=& \frac{m(a^2-3r^2)+r(r^2+a^2)(m'+1)}{a \left [ m+r(m'-1) \right ] } \label{eq:xiregular} \,.
\end{eqnarray}
\end{widetext}
In the above expressions, the ``mass function'' $m(r)$ is defined as $m(r) \equiv f(r)/r$, with $f(r)$ being the function appearing in Eq.~(\ref{eq:rotatingbh}), and with explicit forms of $F(r) \equiv 1-2f(r)/r^2$ for the rotating Bardeen and Hayward BHs given in Eqs.~(\ref{eq:fb},\ref{eq:fh}). Moreover, we have defined $m'(r) \equiv dm(r)/dr$. Finally, while in the Kerr BH case $m(r)=M$, for the rotating Bardeen and Hayward space-times the mass function is given by:
\begin{eqnarray}
m_B(r) &=& M \left ( \frac{r^2}{r^2+q_m^2} \right ) ^{\frac{3}{2}} \label{eq:mrbardeen} \,, \\
m_H(r) &=& M\frac{r^3}{r^3+2M\ell^2} \label{eq:mrhayward} \,.
\end{eqnarray}
Inserting Eqs.~(\ref{eq:mrbardeen},\ref{eq:mrhayward}) into Eqs.~(\ref{eq:eregular},\ref{eq:xiregular}) can be used to determine the relation between $L_z$, ${\cal C}$, and the parameter of closed photon orbits $r$ for the rotating Bardeen and Hayward BHs. Finally, as mentioned in Sec.~\ref{sec:kerr}, the same code we use to compute $R_s(\mu)$ from closed photon orbits is also used to compute the quantity $\varepsilon(\mu,a)$, which is defined in Eq.~(\ref{eq:vareps}) and quantifies the error made in adopting the approximate solution to the B-S quantization condition given by Eq.~(\ref{eq:expansion}). This is given by the following:
\begin{eqnarray}
\varepsilon(\mu,a) = \frac{L}{E} - \frac{\widetilde{L}}{E}\,,
\label{eq:epsilon}
\end{eqnarray}
where $L/E$ is computed by numerically solving the B-S quantization condition given in Eq.~(\ref{eq:BScondition}), whereas $\widetilde{L}/E$ is calculated using Eq.~(\ref{eq:expansionnew}). The value of $\varepsilon(\mu,a)$ is then used to further improve the eikonal QNM-shadow correspondence as per Eq.~(\ref{eq:correspondencenew}).

\bibliography{QNMshadow}

\end{document}